\documentclass[11pt]{article}

\usepackage{float}
\usepackage[colorinlistoftodos,prependcaption,textsize=tiny]{todonotes}
\usepackage{verbatim}
\usepackage{amsmath}
\usepackage{graphicx,psfrag,epsf}
\usepackage{caption}
\usepackage{subcaption}
\usepackage{bm}
\usepackage{enumerate}
\usepackage{amssymb}
\usepackage{hyperref}

\newtheorem{theorem}{Theorem}
\newtheorem{proposition}{Proposition}
\newtheorem{corollary}{Corollary}

\usepackage{dsfont}
\usepackage[lined,boxed]{algorithm2e}

\usepackage{natbib}
\providecommand{\keywords}[1]{\textbf{\textit{Keywords:}} #1}

\def\_[{\ensuremath\left[}
\def\_]{\ensuremath\right]}

\def\_({\ensuremath\left(}
\def\_){\ensuremath\right)}

\DeclareMathOperator{\Var}{Var}

\newcommand{\E}[1]{\mathbb{E} \left[ #1\right] }

\newcommand{\ystar}{y^\star}
\newcommand{\uvec}{\mathbf{u}}
\newcommand{\thetavec}{\theta}
\newcommand{\yvec}{\bm{y}}
\newcommand{\sumnN}{\sum_{n=1}^N}
\newcommand{\sumkN}{\sum_{k=1}^N}

\newcommand{\1}[1]{\mathds{1}\left\{#1\right\} }

\newcommand*{\QED}{\hfill\ensuremath{\square}}

\newcommand{\pball}{\mathbb{P}_{\thetavec}\left( \delta(y,y^\star)\leq \epsilon\right)}
\newcommand{\pballn}{\mathbb{P}_{\thetavec_n}\left( \delta(y,y^\star)\leq \epsilon\right)}
\newcommand{\ind}{\mathds{1}}
\newcommand{\dd}{\mathrm{d}}
\newcommand{\pest}{\hat{L}_\epsilon}
\newcommand{\pestt}{\hat{L}_{\epsilon_t}}

\newcommand{\R}{\mathbb{R}}
\newcommand{\OO}{\mathcal{O}}
\newcommand{\oo}{o}
\newcommand{\Op}{\OO_P}

\newcommand{\EE}{\mathbb{E}}
\newcommand{\Zest}{\hat{Z}_N}
\newcommand{\phiest}{\hat{\phi}_N}
\newcommand{\ubox}{[0,1]^d}
\newcommand{\Unif}{\mathcal{U}}
\newcommand{\eqdef}{:=}
\newcommand{\cvl}{\stackrel{\mathcal{L}}{\rightarrow}}
\newcommand{\Zp}{\mathbb{Z}^+}
\newcommand{\bu}{\mathbf{u}}

\begin{document}

\title{ Improving approximate Bayesian computation via quasi-Monte Carlo}
\author{Alexander Buchholz \qquad Nicolas Chopin \\ 
       ENSAE-CREST}
\date{}

\maketitle

\begin{abstract}
	ABC (approximate Bayesian computation) is a general approach for dealing
	with models with an intractable likelihood. In this work, we derive ABC
	algorithms based on QMC (quasi-Monte Carlo) sequences. We show that the
	resulting ABC estimates have a lower variance than their Monte Carlo
	counter-parts.  We also develop QMC variants of sequential ABC
	algorithms, which progressively adapt the proposal distribution and the
	acceptance threshold. We illustrate our QMC approach through several
	examples taken from the ABC literature.

    The computer code used to perform our numerical experiments is available 
    at \url{https://github.com/alexanderbuchholz/ABC}. 

\end{abstract}

\section{Introduction}
Since its introduction by \cite{Tavare1997} approximate Bayesian computation
(ABC) has received growing attention and has become today a major tool for
Bayesian inference in settings where the likelihood of a statistical model is
intractable but simulations from the model for a given parameter value can be
generated. The approach of ABC is as convincing as intuitive: We first sample a
value from the prior distribution, conditional on this prior simulation an
observation from the model is generated. If the simulated observation is
sufficiently close to the observation that has been observed in nature, we
retain the simulation from the prior distribution and assign it to the set of
posterior simulations. Otherwise the simulation is discarded. We repeat this
procedure until enough samples have been obtained.

Since then several computational extensions related to ABC have been proposed.
For instance the use of MCMC as by \cite{Marjoram2003} has improved the
simulation of ABC posterior samples over the simple accept--reject algorithm.
The use of sequential approaches by \cite{Beaumont2009}, \cite{Sisson2009}, 
\cite{DelMoral2012} and  \cite{Sedki2013} made it possible to exploit the information from
previous iterations and eventually to choose adaptively the schedule of
thresholds $\epsilon$. 
Besides the question of an efficient simulation of high posterior probability
regions, the choice of summary statistics, summarizing the information
contained in the observation and the simulated observation, has been investigated \citep{Fearnhead2012}. 
See \cite{Marin2012} and \cite{Lintusaari2016} for two recent reviews. Moreover,
the introduction of more machine learning driven approaches like random forests
\citep{Marin2016}, Gaussian processes \citep{Wilkinson2014}, Bayesian
optimization \citep{Gutmann2016}, expectation propagation \citep{Barthelme2011}
and neural networks \citep{papamakarios2016fast} have
been proposed. A post-processing approach 
based on nonparametric regression 
was studied in \cite{blum2012approximate}. 

In this paper we take a different perspective and approach the problem of
reducing the variance of ABC estimators. We
achieve this by introducing so called low discrepancy sequences in the
simulation of the proposal distribution. We show that this allows
to reduce significantly the variance of posterior estimates.

The rest of the paper is organized as follows. Section \ref{sec:ABC}
reviews the basic ideas of approximate Bayesian computation and sets the 
notation. Section \ref{sec:QMC} introduces the concept of low
discrepancy sequences. Section \ref{sec:improving_ABC_QMC} brings 
the introduced concepts together and provides the theory that underpins
the proposed idea. Section \ref{sec:non_sequential_applications} presents
a first set of numerical examples. 
Section \ref{sec:sequential_sampling} explains how to
use our ideas in a sequential procedure which adapts progressively the 
proposal distribution and the value of $\epsilon$. 
Section \ref{sec:sequential_applications} illustrates the resulting sequential ABC
procedure. Section \ref{sec:discussion} concludes.

\section{Approximate Bayesian computation} \label{sec:ABC}

\subsection{Reject-ABC} 

Approximate Bayesian computation is motivated by models such that (a) 
the likelihood function is difficult or expensive to compute; 
(b) simulating from the model (for a given parameter $\thetavec$) is feasible. 

The most basic ABC algorithm is called reject-ABC. It consists in simulating pairs $(\thetavec, y)$, 
from the prior $p(\thetavec)$ and the likelihood $p(y|\thetavec)$, and keeping those pairs such that 
$\delta(y,y^\star)\leq \epsilon$, where $y^\star$ is the actual data, 
and 
$\delta: \mathcal{Y} \times \mathcal{Y}
\rightarrow  \mathbb{R}^+$ is some distance (e.g. Euclidean). 
This is done until $N$ pairs are accepted. 
The target density of this rejection algorithm is: 
\begin{eqnarray*}
	p_{\epsilon}(\thetavec, y) = \frac{1}{Z_\epsilon}
p(\thetavec) p(y|\thetavec) \1{\delta(y, \ystar) \leq \epsilon},
\end{eqnarray*}
and its marginal density with respect to $\thetavec$ is: 
\begin{equation}\label{eq:abc_target_marginal}
	p_\epsilon(\thetavec) = \frac{1}{Z_\epsilon} p(\thetavec)\pball 
\end{equation} 
where $\mathbb{P}_{\thetavec}$ denotes a probability with respect to $y\sim p(y|\thetavec)$, 
and $Z_\epsilon = \int_\Theta p(\thetavec)\pball \mathrm{d} \thetavec$ is the normalising constant.

As $\epsilon\rightarrow 0$, \eqref{eq:abc_target_marginal} converges to the true posterior density. 
Actually, $\delta$ is often not a distance but a pseudo-distance of the form: 
$\delta(y, y^\star) = \| s(y) - s(y^\star) \|_2$, where $\|\cdot \|_2$ is the Euclidean norm, and 
$s(y)$ is a low-dimensional, imperfect summary of $y$. In that case, 
$p_\epsilon(\thetavec) \rightarrow p(\thetavec|s(y^\star))$. This introduces an extra level 
of approximation, which is hard to assess theoretically and practically. However, in this 
paper we focus on how to approximate well \eqref{eq:abc_target_marginal} for a given $\delta$
(and  $\epsilon$), 
and we refer to e.g. \cite{Fearnhead2012} 
for more discussion on the choice of $\delta$ or $s$. 

\subsection{Pseudo-marginal importance sampling}\label{sub:pseudo}

A simple generalisation of reject-ABC is described in Algorithm \ref{algo:abc_is}. 
For $n=1,\ldots, N$, we 
sample the parameter $\thetavec_n\sim q(\thetavec)$, the latent variable $x_n\sim q_{\thetavec_n}(x)$, 
and reweight $(\thetavec_n, x_n)$ according to 
\[
	w_n = \frac{p(\thetavec_n)}{q(\thetavec_n)} \times \pest(x_n)  
\]
where, for $x\sim q_\theta$,  $\pest(x)$ is an unbiased estimate of the probability $\pball$:  
\[\int q_\theta(x) \pest(x) \dd x = \pball . 
\]

\begin{algorithm}[H] \label{algo:abc_is}
\SetKwInOut{Input}{Input}
 \Input{Observed $\ystar$, prior distribution $p(\thetavec)$, proposal distribution
 $q(\thetavec)$, simulator $ q_{\thetavec_n}(x)$, distance function $\delta(\cdot, \cdot)$, target
 threshold $\epsilon$, number of simulations $N$} 
 \KwResult{Set of weighted samples $(\thetavec_n, x_n, w_n)_{n \in 1:N}$}
 \For{$n = 1$ to $N$}{
  Sample $\thetavec_n \sim q(\thetavec)$ \\
  Sample $x_n \sim q_{\thetavec_n}(x)$\\
  Set $w_n = p(\thetavec_n) \pest(x_n) / q(\thetavec_n)$ 
 }
 \caption[ABC_accept_reject]{ABC importance sampling algorithm}
\end{algorithm}

The marginal density (with respect to $\thetavec$) of the target density 
of this importance sampling scheme is again \eqref{eq:abc_target_marginal}. 
In particular, the quantity 
\begin{eqnarray}\label{eq:is_estimator}
\hat{\phi}_N = \frac{\sum_{n=1}^N w_n \phi(\thetavec_n) }{\sum_{n=1}^N w_n},
\end{eqnarray}
is a consistent (as $N\rightarrow \infty$ and under appropriate conditions) estimate of expectation 
$\mathbb{E}_{p_\epsilon(\thetavec)}[\phi(\thetavec)]$, for $\phi:\Theta \rightarrow \R$.  
Since the importance weight involves an unbiased estimator, the whole procedure may be viewed 
as a pseudo-marginal sampler, in the spirit of \cite{Andrieu2009}. 

A a special case, take the proposal $q(\thetavec)$ to be equal to the prior, $p(\thetavec)$, 
and take $x=y$, $\pest(x) = \ind\left\{ \delta(y, y^\star) \leq \epsilon \right\}$; 
then we recover essentially the same procedure as reject-ABC (except that $N$ stands 
for the number of proposed points, rather than the number of accepted points). 
However, the generalized scheme allows us (a) to sample $\thetavec_n$ from a distribution $q(\thetavec)$  
which may be more likely (than the prior) to generate high values for the probability $\pball$; 
and (b) to use a more sophisticated unbiased estimate for $\pball$. 

Regarding (b), we consider two unbiased schemes in this work. In the first part, we focus on:  
\begin{eqnarray} \label{eq:weight_estimator_several_m}
 x=y_{1:M}, \quad 
 q_{\thetavec}(x)=\prod_{m=1}^M p(y_m|\thetavec),\quad 
 \pest(x) = \frac 1 M \sum_{m=1}^M \ind\{ \delta(y_m,y^\star) \leq \epsilon \}. 
\end{eqnarray}
for a certain $M \geq 1$. The possibility to associate more than one datapoints to each 
parameter $\thetavec_n$ was considered in e.g. \cite{DelMoral2012}. \cite{Bornn2015} showed that $M=1$ usually 
represents the best variance vs CPU time trade-off when using Monte Carlo sampling, 
however we shall see that this result does not hold when using QMC. 

Later on in the paper, we shall consider an alternative unbiased estimator, based on 
properties of the negative binomial distribution. More precisely, assume that, for 
a given $\thetavec$, we sample sequentially $y_1, y_2, \ldots \sim p(y|\thetavec)$, until 
we reach the time $k$ where $r\geq 2$ datapoints are such that $\delta(y_n, y^\star)\leq \epsilon$; 
then $k$ is distributed according to a negative
binomial distribution with parameters $r$ and $p=\pball$, and the minimum-variance unbiased estimator of 
$\pball$ is \cite[Chap. 8]{johnson2005univariate}: 
\[
	\pest(x) = \frac{r - 1}{k - 1}
\]
where $x=y_{1:k}$. 

The second unbiased estimator is closely related, but not equivalent to, the $r$-hit kernel 
of \cite{Lee2012}; see also \cite{LeeLatuszynski}. 
Specifically, \cite{Lee2012} proposed an MCMC kernel that generates \emph{two} 
negative binomial variates (one for the current point, and one for the proposed point) at each iteration. 
The invariant distribution of this kernel is such that, marginally, $\thetavec$ is distributed according to 
\eqref{eq:abc_target_marginal}.

In more practical terms, we shall use the latter estimator in situations where we would like to 
set $\epsilon$ beforehand to some value such that $\pball$ may be small. In that case, 
this estimator automatically adjusts the CPU budget (i.e. the number of simulations from the likelihood) 
so as to ensure that the number of simulated $y-$values is non-zero. 
But we shall return to this point in Section \ref{sec:sequential_sampling}.

\section{Quasi-Monte Carlo} \label{sec:QMC}

\subsection{QMC Overview}%
\label{sub:qmc_overview}

This section gives a brief overview of QMC and the underlying theory; for a
more in-depth presentation, see e.g.  the book of \cite{Lemieux_book}, the book
of \cite{leobacher2014introduction} or Chapter 5 in \cite{glasserman2013monte}. 

QMC sequences (also called low discrepancy sequences), are used
to approximate integrals over the $[0,1]^d$ hypercube:
\begin{eqnarray*}
\E{\psi(U)} = \int_{[0,1]^d} \psi(u) \dd u, 
\end{eqnarray*}
that is the expectation of the random variable $ \psi(U)$, where $ U \sim
\mathcal{U} \left([0,1]^d \right)$. The basic Monte Carlo approximation of the
integral is  $\hat{I}_N := N^{-1}
\sumnN \psi(\uvec_n)$, where each $\uvec_n\sim \mathcal{U} \left([0,1]^d \right)$.
The error of this approximation is $\Op(N^{-1/2})$, since
$\Var[\hat{I}_N] = \Var[\psi(U)]/N$.

\begin{figure}[h]
	\center{
   \includegraphics[scale=0.5]{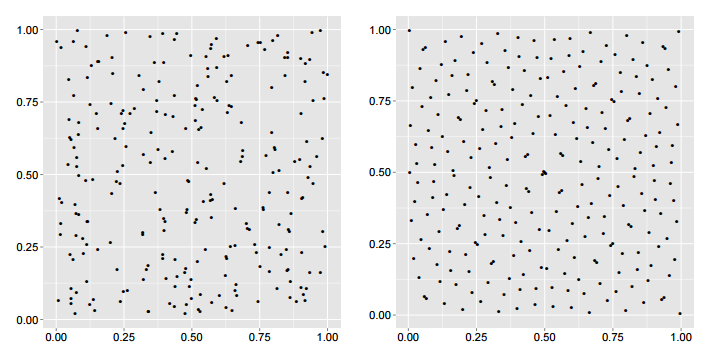}
   \caption{\label{fig:sequence} Uniform random (left) and QMC (right) point
   sets of length 256 in $[0,1]^2$. The QMC sequence covers the target space 
   more evenly than the random uniform sequence.}
   }
\end{figure}

It is possible to improve on this basic approximation, by replacing the random
variables $\uvec_n$ by a low-discrepency sequence; that is, informally, a
deterministic sequence that covers $[0,1]^d$ more regularly. This idea is
illustrated in Figure \ref{fig:sequence}.

More formally, the general notion of discrepancy of a given sequence is defined
as follows:
\begin{eqnarray*}
D(\uvec_{1:N}, \mathcal{A}) := \sup_{A \in \mathcal{A}} \left| \frac{1}{N} \sumnN \1{\uvec_n \in A }  - \lambda_d(A)\right|,
\end{eqnarray*}
where $\lambda_d(A)$ is the volume (Lesbegue measure on $\mathbb{R}^d$) of $A$
and $\mathcal{A}$ is a set of measurable sets.  When we fix the sets $A$ to be
intervals anchored at $0$ we obtain the so called star discrepancy:
\begin{eqnarray*}
D^*(\uvec_{1:N}) := \sup_{[0,\mathbf{b}] } \left| \frac{1}{N} \sumnN \1{\uvec_n
\in [0, \mathbf{b}] }  - \prod_{i=1}^d b_i\right|,
\end{eqnarray*}
where $[0,\mathbf{b}]= \prod_{i=1}^d [0,b_i], 0 \leq b_i \leq 1$. The
importance of the notion of discrepancy and in particular the star discrepancy
is highlighted by the Koksma-Hlawka inequality \citep{doi:10.1002/0471667196.ess4085.pub2}, which relates the error
of the integration to the coverage of the space and the variation of the
function that is integrated:
\begin{eqnarray*}
\left| \int_{[0,1]^d} \psi(\uvec) \dd\uvec -  \frac{1}{N} \sumnN \psi(\uvec_n) \right| \leq V(\psi) D^*(\uvec_{1:N}),
\end{eqnarray*}
where $V(\psi)$ is the variation in the sense of Hardy and Krause \citep{zbMATH02650645}.
The actual definition of this quantity is a bit involved, but essentially it
measures in some way the smoothness of the function $\psi$;
see \cite{kuipers2012uniform} and \cite{leobacher2014introduction} for more details.

It is possible to construct sequences $\uvec_n$ such that, when $N$ is fixed in
advance, $D^*(\uvec_{1:N})$ is $\mathcal{O}\left( N^{-1}(\log N)^{d-1} \right)$,
and, when $N$ is allowed to grow, i.e., the sequence must be generated
iteratively, then $D^*(\uvec_{1:N}) = \mathcal{O}\left( N^{-1}(\log N)^{d}
\right)$. Then $\forall \tau > 0$ the error rate is
$\mathcal{O}\left(N^{-1+\tau} \right)$.
Consequently, QMC integration schemes are asymptotically more efficient than MC
schemes.
One observes in practice that QMC integration outperforms MC integration even
for small $N$ in most applications, see e.g. the examples in Chapter 5 of
\cite{glasserman2013monte}.

\subsection{Randomized quasi-Monte Carlo}
A drawback of QMC is that it does not come with an easy way
to assess the approximation error.
RQMC (randomized quasi-Monte Carlo) amounts to introduce randomness in
a QMC sequence, in such a way that
$\uvec_n \sim \mathcal{U}\left([0,1]^d\right)$, marginally. The quantity
$\hat{I}_N = N^{-1} \sum_{n=1}^N \psi(\uvec_i)$ is then an unbiased estimate
of the integral of interest. One may assess the approximation
error by computing the empirical variance over repeated simulations.

The simplest way to obtain an RQMC sequence is to randomly shift a QMC sequence:
Let $\mathbf{v} \sim \mathcal{U}\left([0,1]^d\right)$, and $\uvec_{1:N}$ a QMC sequence; then
\begin{eqnarray*}
	\hat{\uvec}_n :=  \uvec_n + \mathbf{v} \mod 1\, \mbox{(component wise)}
\end{eqnarray*}
is an RQMC sequence.

A more sophisticated approach, called scrambled nets, was introduced by
\cite{owen_scrambled_1997} and later refined in \cite{owen2008local}.  The main
advantage of this approach is that under the assumption of smoothness of the
derivatives of the function, the speed of convergence can be even further
improved, as stated in the following Theorem.  \begin{theorem}
    \label{theorem:owen} \citep{owen2008local} Let $f:[0,1]^d\rightarrow
    \mathbb{R}$ be a function such that its cross partial derivatives up to
    order $d$ exist and are continuous, and let $(\uvec_n)_{n \in 1:N}$ be a
    relaxed scrambled $(\lambda, t, m, d)$-net in base $b$ with dimension $d$
    with uniformly bounded gain coefficients.  Then, \[\Var\left(\frac{1}{N}
\sumnN f(\uvec_n) \right) = \mathcal{O}\left(N^{-3} \log(N)^{(d-1)}\right) , \]
where $N=\lambda b^m$.  \end{theorem} In words, $\forall \tau > 0$ the RQMC
error rate is $\OO(N^{-3/2+\tau})$ when a scrambled $(\lambda, t, m, d)$-net is
used. This result has the only inconvenience that the rate of convergence only
holds for certain $N$.  However, a more general result has recently been shown
by \cite{gerber2015integration}[Corollary 1], where if $f \in L^2$ and
$(\uvec_n)_{n \in 1:N}$ is a scrambled $(t, d)$-sequence, then $\forall N \in
\Zp,$ 
\[ \Var\left(\frac{1}{N} \sumnN f(\uvec_n) \right) = \oo
\left(N^{-1}\right).  
\]

The construction of scrambled nets and sequences is quite involved. As the focus of our paper is the application 
and not the construction of these sequences, we refer the reader for more details to \citet{l2016randomized} or \cite{dick2013high}. 
In the following, when speaking about an RQMC sequence, we will assume that this sequence is a 
scrambled $(t, d)$-sequence. 

\subsection{Mixed sequences and a central limit theorem}

One drawback of low discrepancy sequences is that the speed
of convergence deteriorates with the dimension.
In some situations, a small number of components contributes significantly
to the variance of the target. One then might choose to use a
low discrepancy sequence for those components and an ordinary Monte Carlo
approach for the rest. This idea of using a \emph{mixed sequence}
is closely linked to the concept of
effective dimension, see \cite{Owen1998}.
Based on the randomness induced by the Monte Carlo part a central limit
theorem (CLT) may be established:
\begin{theorem} \label{theorem:mvt_clt_mixed}
\citep{okten2006central} Let $u_k = (q^{1:d}_k, X^{d+1:s}_k)$ be a mixed
sequence of dimension $s$ where $q^{1:d}_k$ denotes the deterministic QMC part
and $X^{d+1:s}_k$ denotes the random independent MC part.
Let $f:[0,1]^s \rightarrow \mathbb{R}^t, t \in \Zp$ a bounded,
square integrable function, $Y_k=f(u_k)$,
$\hat{I}_N = N^{-1} \sumkN Y_k$, and
\begin{align*}
	\mu_k & \eqdef  \mathbb{E}[Y_k] =  \int_{[0,1]^{s-d}} f(u_k) \dd X^{d+1:s},\\
	S_N & \eqdef \frac{1}{N} \left( \sumkN Y_k - \sumkN \mu_k \right)
	=  \left( \hat{I}_N - \frac{1}{N} \sumkN \mu_k \right),\\
  \sigma_k^2 & \eqdef  \Var [Y_k] = \int_{[0,1]^{s-d}} f(u_k)f(u_k)^T \dd X^{d+1:s} \\
		    & \qquad - \left( \int_{[0,1]^{s-d}} f(u_k) \dd X^{d+1:s}
		    \right) \left( \int_{[0,1]^{s-d}} f(u_k) \dd X^{d+1:s}
	    \right)^T, \\
  C_N^2 & \eqdef \Var [N \hat{I}_N]  =  \sumkN \sigma_k^2.
\end{align*}
Then, as $N\rightarrow +\infty$,  $C_N^2/ N \rightarrow C_{\mathrm{qmc-mixed}}^2$ and
\begin{eqnarray*}
	N^{1/2} S_N  \stackrel{\mathcal{L}}{\rightarrow}
	\mathcal{N}\left(0,C_{\mathrm{qmc-mixed}}^2\right),
\end{eqnarray*}
where
\begin{align*}
C_{\mathrm{qmc-mixed}}^2 =
&  \int_{[0,1]^{s}} f(x)f(x)^T \dd x  \\
& - \int_{[0,1]^{d}}\left( \int_{[0,1]^{s-d}} f(u) \dd X^{d+1:s} \right) \left(
\int_{[0,1]^{s-d}} f(u) \dd X^{d+1:s} \right)^T \dd q^{1:d}.
\end{align*}
\end{theorem}
As a direct corollary of the previous Theorem we obtain that,
provided $f$ has a finite variation in the sense of Hardy and Krause,
$	N^{1/2} (\hat{I}_N - I)    \stackrel{\mathcal{L}}{\rightarrow}
\mathcal{N}(0,C_{\mathrm{qmc-mixed}}^2)$,
where $I = \int f(u) \dd u$. This is due to the fact that
\[ N^{1/2} \left(\hat{I}_N - I\right) =
N^{1/2} S_N +
 N^{1/2} \left( \frac 1 N \sumkN \mu_k - I \right)
\] and the second term on the right hand side
converges deterministically to $0$.
\cite{okten2006central} present only a univariate version of
their central limit theorem; the extension to the multivariate case is 
straightforward. 

Moreover, their work shows that the asymptotic variance of the mixed
sequence estimator is smaller than for the same estimator based on Monte Carlo
sequences in dimension one. We extend this result to the multivariate case.

\begin{corollary} \label{corl:asymptotic_variance_reduction}
Let $C_{\mathrm{qmc-mixed}}^2$ be the asymptotic variance of an estimator based
on a mixed sequence as defined in Theorem \ref{theorem:mvt_clt_mixed}. Let
$C_{\mathrm{mc}}^2$ be the variance of the same estimator based on a pure MC
sequence, e.g., when $d=0$. Then
\begin{eqnarray*}
	C_{\mathrm{qmc-mixed}}^2 \preceq C_{\mathrm{mc}}^2
\end{eqnarray*}
in the sense of positive definite matrices.
\end{corollary}
Moreover, we present a result here that allows us to apply the same technique
to mixed sequences that combine Monte Carlo and randomized quasi-Monte Carlo
sequences.
\begin{theorem} \label{theorem:mvt_clt_mixed_rqmc}
Let $S_N^{RQMC}$ be the MC-RQMC equivalent of $S_N$ under the same conditions
as in Theorem \ref{theorem:mvt_clt_mixed}. Then
	\begin{eqnarray*}
		 N^{1/2} S_N^{\mathrm{RQMC}}
		 \cvl
		 \mathcal{N}(0,C_{\mathrm{rqmc-mixed}}^2),
	\end{eqnarray*}
	where $C_{\mathrm{rqmc-mixed}}^2 = C_{\mathrm{qmc-mixed}}^2$.
\end{theorem}
These results may be understood as follows. The randomness in the Monte Carlo
sequence allows the construction of a central limit theorem.  The part
associated to the (R)QMC sequences converges faster to zero than the part
associated to the Monte Carlo sequence. This leads to a reduced asymptotic variance
for estimators based on mixed sequences.

\section{Improved ABC via (R)QMC} \label{sec:improving_ABC_QMC}

Recall that we described our ABC importance sampler as an algorithm 
that samples pairs $(\thetavec_n, x_n)$ from $q(\thetavec)q_{\thetavec}(x)$, 
where $x_n$ consists of datapoints generated from the model. In most ABC problems, 
using (R)QMC to generate 
the $\thetavec_n$ should be easy, but this should not be the case for the $x_n$'s. 
Indeed, the simulator used to generate datapoints from the model may be a complex 
black box, which may require a very large, or random, number of uniform variates. 
Thus, we contemplate from now on generating the $\thetavec_n$'s using (R)QMC. That is, 
$\thetavec_n = \Gamma(\uvec_n)$, where $\uvec_{1:N}$ is a QMC or RQMC sequence,
and 
$\Gamma$ is a function such that $\Gamma(U)$, $U\sim \Unif\left(\ubox\right)$, is
distributed 
according to the proposal $q(\thetavec)$; and $x_n|\thetavec_n \sim
q_{\thetavec_n}$ 
is a random variate. In other words, $(\thetavec_n, x_n)$ is a mixed sequence. 

We already know from the previous section that an estimate based on a mixed sequence
converges at the Monte Carlo rate, $\Op(N^{-1/2})$, but has a smaller asymptotic variance
than the same estimate based on Monte Carlo. In fact, a similar result may be established 
directly for the actual variance. 
Let $\hat{I}_N := \sum_{n=1}^N \varphi(\thetavec_n, x_n)/N$ be an empirical average 
for some measurable function $\varphi$. 
For simplicity, we assume here that the 
$\thetavec_n$'s are either random variates, or RQMC variates. 
That is, in both cases, $\thetavec_n\sim q$ marginally.
Then 
\begin{align}
\Var[\hat{I}_N] 
& = \Var\left[\EE\{\hat{I}_N | \thetavec_{1:N} \} \right]
+ \EE\left[ \Var\{ \hat{I}_N | \thetavec_{1:N}\} \right]  \nonumber\\
& = \Var\left[ \frac 1 N \sum_{n=1}^N \EE_{x_n\sim q_{\thetavec_n}}
\left\{ \varphi(\thetavec_n, x_n)|\thetavec_n) \right\} \right]
+ \frac 1 N \times  \EE_{\thetavec_n\sim q}\left[ \Var_{x_n \sim q_{\thetavec_n}} 
\left\{\varphi(\thetavec_n,x_n)|\thetavec\right\} \right] 
\label{eq:var_decomposition}
\end{align}

The first term is $\OO(N^{-1})$ when the $\thetavec_n$'s are generated using
Monte Carlo, and should be $\oo(N^{-1})$ under appropriate conditions when the
$\thetavec_n$'s are an RQMC sequence.  On the other hand, the second term is
$\OO(N^{-1})$ in both cases. As a corrolary, the variance of $\hat{I}_N$ is
smaller when using a mixed sequence, for $N$ large enough. 

The point of the following sections is to generalize this basic result to
various ABC estimates of interest.

\subsection{Improved estimation of the normalization constant}
\label{sub:improved_normconstant}

We first consider the approximation of the normalization constant of the ABC
posterior: 
\begin{eqnarray*}
  Z_\epsilon =
  \int \mathbb{P}_{\thetavec} \left( \delta(y, \ystar) \leq \epsilon \right)
  p(\thetavec) \dd \thetavec 
  = \int   \pest(x) q_{\thetavec}(x) p(\thetavec)\dd x \dd \thetavec.
\end{eqnarray*}
Recall that, for the moment, we take $x=y_{1:M}$, $q_{\thetavec}(x)=
\prod_{m=1}^M p(y_m|\thetavec)$ and  
$$\pest(x) = \frac 1 M \sum_{m=1}^M \ind\left\{ \delta(y_m, y^\star)\leq \epsilon \right\} .$$ 
Thus, a natural estimator of $Z_\epsilon$  is 
\begin{eqnarray}
	\Zest \eqdef \frac{1}{N} \sum_{n=1}^N \frac{p(\thetavec_n)}{q(\thetavec_n)} 
	\left[\frac{1}{M} \sum_{m=1}^M \ind\left\{ \delta(y_{n,m}, y^\star) \leq \epsilon\right\}\right]
\end{eqnarray}
where the $\thetavec_n$'s are either a Monte Carlo or RQMC sample from the
proposal $q(\thetavec)$, and $y_{n,m}\sim p(y|\thetavec_n)$ for $n=1,\ldots,N$,
$m=1,\ldots, M$. 

When the $\thetavec_n$'s are a Monte Carlo sample, it is always best to take
$M=1$, as noted by \cite{Bornn2015}.
This may be seen by calculating both terms of decomposition 
\eqref{eq:var_decomposition} when applied to the estimator of the normalization constant $\Zest$: 
\begin{align}
 \Var\left[\EE\{\Zest | \thetavec_{1:N} \} \right] 
 & = \frac 1 N \times \Var_q\left[ \frac{p(\thetavec)}{q(\thetavec)} \pball \right] \label{eq:varE}\\ 
 \EE\left[ \Var\{ \Zest | \thetavec_{1:N}\} \right]  
 & =	\frac 1 {NM} \times \int_\Theta \frac{p(\thetavec)^2}{q(\thetavec)} \pball \{1 -\pball\} \dd \thetavec .\label{eq:Evar}
\end{align}
Increasing $M$ increases the CPU cost and decreases the variance of $\Zest$. To
account for both simultaneously, we look at the adjusted variance, $M\times
\Var\left[ \Zest\right]$. From \eqref{eq:varE} and \eqref{eq:Evar}, 
we see that the adjusted variance increases with $M$, hence 
the best CPU time vs error trade-off is obtained by taking $M=1$. 

Now, consider the situation where the $\thetavec_n$'s form an RQMC sequence. 
As noted in the previous section, \eqref{eq:Evar} still holds due to the unbiasedness property of RQMC sequences, however the first \eqref{eq:varE} term of the 
decomposition should converge faster. 

\begin{proposition} \label{theorem:decomp_variance_norm_constant_rqmc}
	Let $f(\thetavec) = \{ p(\thetavec) / q(\thetavec) \} 
	\pball $, assume that $\thetavec_n=\Gamma(\uvec_n)$ where $\uvec_{1:N}$
	is a scrambled $(\lambda, t, m, d)$-net, and assume that  $f\circ \Gamma \in L^2$ . Then, 
\[\Var\left[\EE\{ \Zest | \thetavec_{1:N} \} \right] = \oo\left( N^{-1}\right). 
\]
\end{proposition}

This result is a direct consequence of Corollary 1 of \cite{gerber2015integration} and the fact
\[\EE\{\Zest|\thetavec_{1:N}\} 
	= \frac 1 N \sum_{n=1}^N f(\thetavec_n)
	= \frac 1 N \sum_{n=1}^N f\circ\Gamma(\uvec_n)
. \]
It has two corollaries. First, the variance of $\Zest$ is smaller when using a
RQMC sequence for the $\thetavec_n$'s (for $N$ large enough). Second, in that
case, the adjusted variance is such that 
$M\Var[\Zest] = \OO(N^{-1})$, with a constant that does not depend on $M$. 
Thus taking $M>1$ (within a reasonable range) should have basically no impact
on the CPU time vs error trade-off in the RQMC case. 

Taking $M>1$ has the following advantage: it makes it possible to consistently
estimate \eqref{eq:Evar} with the quantity 
\begin{eqnarray} \label{eq:estimator_var_normalizing_constant}
	\hat{\sigma}^2(Z_\epsilon) \eqdef 
	\frac 1 {N^2(M-1)} \times \sum_{n=1}^N \frac{p(\thetavec_n)^2}{q(\thetavec_n)^2}
	\pest(x_n) \{1 - \pest(x_n)\}. 
\end{eqnarray}
where  
$\pest(x_n) = M^{-1} \sum_{m=1}^M \ind\{ \delta(y_{n,m}, y^\star) \leq \epsilon \}$. 
As \eqref{eq:Evar} corresponds to the non-negligible part of the variance of
$\Zest$, this allows us to obtain asymptotic confidence intervals for $\Zest$. 

We have focused on the RQMC case for now on, but a similar result holds for QMC
sequences.  Note, however, that we cannot use directly decomposition
\eqref{eq:var_decomposition} when the $\thetavec_n$'s are deterministic. 

\begin{proposition} \label{theorem:decomp_variance_norm_constant_qmc}
    Assume that $\bu_{1:N}$ is a deterministic low-discrepancy sequence, 
    that $f\circ \Gamma$ (where $f$ and $\Gamma$ are defined as in 
	Proposition \ref{theorem:decomp_variance_norm_constant_rqmc}) has a
	finite variation in the sense 
	of Hardy and Krause, and that the ratio $p/q$ is upper-bounded, 
$p(\thetavec)/q(\thetavec)\leq C$, then 
$$ M \times \EE \left[ \left(\Zest - Z_\epsilon \right)^2 \right] = \OO(N^{-{1}}) $$
with a constant that does not depend on $M$. Furthermore, the mean square 
error above is smaller than in the Monte Carlo case, for $N$ large enough. 
\end{proposition}

\subsection{Improved estimation of general importance sampling estimators}

We now turn to the analysis of general importance sampling estimators of the form 
\begin{eqnarray} \label{eq:importance_sampling}
	\phiest = \frac{\sum_{n=1}^N w_n \phi(\thetavec_n) }{\sum_{n=1}^N w_n} .
\end{eqnarray}

As these estimators are ratios, we cannot apply decomposition \eqref{eq:var_decomposition} directly. 
However, we may apply the following inequality, due to \cite{agapiou2015importance}: 
\[
	\EE \left\{ \phiest - \EE_{p_\epsilon} \phi \right\}^2 \leq 
	\frac{2}{Z_\epsilon^2} \left( 
		\EE\left\{ \frac 1 N \sum_{n=1}^N w_n \phi(\thetavec_n) - Z_\epsilon \EE_{p_\epsilon} \phi(\thetavec)\right\}^2 
		+ \EE\left\{ \frac 1 N \sum_{n=1}^N w_n - Z_\epsilon \right\}^2 
	\right)
\]
provided $|\phi| \leq 1$. 
Both terms are mean squared errors of empirical averages, and hence may be bounded directly 
using a decomposition of variance and the results of the previous section. 
Thus, we see that, again, when the $\thetavec_n$ 
are generated with (R)QMC, the mean squared error of estimate $\phiest$ is 
$\OO(M^{-1}N^{-1})$ as $N\rightarrow +\infty$. 
However, this inequality does not make it possible to compare the performance of our 
RQMC-ABC procedure with Monte Carlo-based ABC. For this, we now consider the asymptotic behavior
of these estimators.

\begin{theorem} \label{theorem:clt_importance_sampling}
Let $\phi:\Theta \rightarrow \R$ be a bounded function, 
$\bar{\phi} = \phi - \EE_{p_\epsilon}\phi$, 
$\phiest$ defined as \eqref{eq:importance_sampling}, then, 
under the same conditions as Proposition \ref{theorem:decomp_variance_norm_constant_qmc}, 
and assuming further that function 
$\uvec \rightarrow \bar{\phi}(\Gamma(\uvec)) f(\Gamma(\uvec))$ has a finite 
variation (in the sense of Hardy and Krause), one has that 
\begin{equation*} \label{eq:tcl_conv_unif}
	N^{1/2} \left( \phiest -  \EE_{p_\epsilon}\phi \right) 
	\cvl \mathcal{N}\left(0,\sigma_{\text{mixed}}^2(\phi)\right),
\end{equation*} 
where, using the short-hand $b(\thetavec)=\pball$, 
\begin{equation}\label{eq:clt_variance} 
	\sigma_\mathrm{mixed}^2(\phi) = 
	\frac{1}{MZ_\epsilon^2} 
	\int_\Theta \frac{ p(\thetavec)^2 }{q(\thetavec)} \bar{\phi}(\thetavec)^2 
	b(\thetavec) \{1 - b(\thetavec)\}
	\dd\thetavec . 
\end{equation} 
\end{theorem} 

Alternatively, if the parameter values $\thetavec_n$ were generated through 
Monte Carlo sampling, one would obtain a similar central limit theorem, 
but with asymptotic variance 
\[
	\sigma_\mathrm{MC}^2(\phi) = 
	\frac{1}{Z_\epsilon^2} 
	\int_\Theta \frac{ p(\thetavec)^2 }{q(\thetavec)} \bar{\phi}(\thetavec)^2 
	\left[ 	\frac{b(\thetavec) \{1 - b(\thetavec)\}}{M} 
+ b(\thetavec)^2 \right]
	\dd\thetavec
\]
which is larger than or equal to $\sigma_\mathrm{mixed}^2(\phi)$. 

It is possible to obtain a similar result for  RQMC sequences by 
using Theorem \ref{theorem:mvt_clt_mixed_rqmc}.

As for the normalising constant, we observe that the adjusted (asymptotic)
variance, i.e. $M \times \sigma_\mathrm{mixed}^2(\phi)$, is constant with respect
to $M$. Thus, taking $M>1$ does not deteriorate the performance of the algorithm 
(in terms of variance relative to CPU time). And it makes it possible to estimate
consistently the asymptotic variance 
\eqref{eq:clt_variance} (and thus compute confidence intervals) using 
\[
	\widehat{\sigma}_\mathrm{mixed}^2(\phi) = 
	\frac{1}{(\Zest)^2 N(M-1)} 	
	\sum_{n=1}^N \frac{p(\thetavec_n)^2}{q(\thetavec_n)}
	\{\phi(\thetavec_n) - \phiest\}^2 \hat{L}_\epsilon(x_n) 
	\left\{1 - \hat{L}_\epsilon(x_n) \right\}. 
\]

\section{Numerical examples} 
\label{sec:non_sequential_applications} 

We illustrate in this section the improvement brought by (R)QMC through several
numerical examples.  Code for reproducing the results of this section and of
Section \ref{sec:sequential_applications} is available at
\url{https://github.com/alexanderbuchholz/ABC}.

Thus we compare three different approaches, all corresponding to Algorithm 
\ref{algo:abc_is}, but with particles generated using either Monte Carlo
(ABC-IS), Quasi-Monte Carlo (ABC-QMC), or randomised QMC (ABC-RQMC). 
For the generation of the (R)QMC sequences we use the \texttt{R} package
\texttt{randtoolbox} \citep{randtoolbox} and generate Sobol sequences (QMC), 
or Owen-type scrambled Sobol sequences (RQMC), see \cite{Owen1998}. 

We take $q(\thetavec)=p(\thetavec)$, i.e. points are generated from the prior, 
and, unless explicitely stated, we take $M=1$. (The problem of adaptively choosing 
$q$ will be considered in the next section.) 

In this case, weights $w_n$ are either 0 or 1 (according to whether $\delta(y_n,
y^\star)\leq \epsilon$), and we set $\epsilon$ so that the proportion of 
non-zero weights is close to some pre-specified value, e.g. $10^{-3}$. 

\subsection{Toy model} \label{ssec:toy_model_non_sequential}

The first model we consider is the toy model used in \cite{Marin2012} that
tries to recover the mean of a superposition of two Gaussian distributions with
identical mean and different variances: 
\begin{eqnarray*}
\thetavec &\sim& \mathcal{U}\left([-10,10]^d\right) , \\
y|\thetavec &\sim & \frac{1}{2} \mathcal{N}(\thetavec; 0.1 I_d) + \frac{1}{2} \mathcal{N}(\thetavec; 0.001 I_d).
\end{eqnarray*}
The use of this model is motivated by the fact that the dimension of the model
$d$ can be scaled up easily. We set $\ystar = 0_d$ and
$\delta(y,\ystar) = \| y-\ystar \|_2$. 
Posterior density \eqref{eq:abc_target_marginal} 
may be calculated exactly in this particular case. 
(The resulting expression depends on the cdfs 
of non-central $\chi^2$ distributions.)

\begin{figure}[H]
\centering
\begin{subfigure}{.45\textwidth}
  \centering
  \includegraphics[width=1\linewidth]{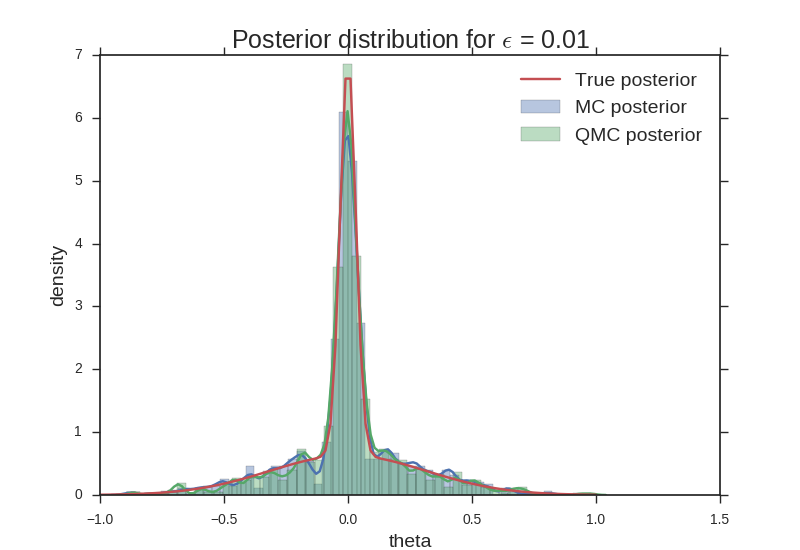}
\caption{}
  \label{fig:mixed_gaussian_posterior}
\end{subfigure}%
\begin{subfigure}{.45\textwidth}
  \centering
  \includegraphics[width=1\linewidth]{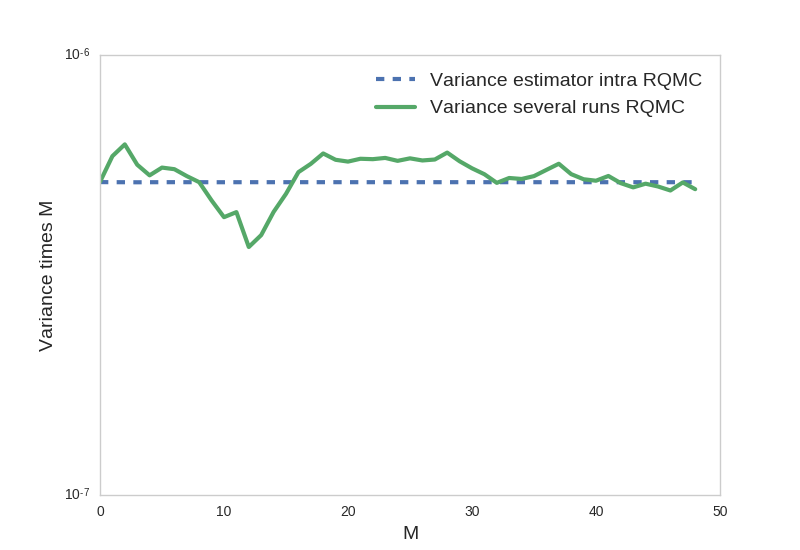}
\caption{}
\label{fig:variance_during_after_several_m}
\end{subfigure}
\caption{Left: Kernel density estimation of the approximation of the 
	posterior distribution based on $N=10^6$ simulations and the
threshold $\epsilon = 0.01$ for $d=1$. The exact posterior can be calculated analytically.
The approaches based on MC and QMC essentially recover the same distribution.
Right: Adjusted variance (variance times $M$) of 
the normalization constant as a function of $M$: the dashed line corresponds to 
the variance estimator given by \eqref{eq:estimator_var_normalizing_constant},
the solid line corresponds to the empirical variance of the estimator based on 75 runs. 
The results are based on $N=10^5$ simulations, $\epsilon=1$, $d=1$, and an RQMC sequence for 
the $\thetavec_n$'s. The adjusted variance stays roughly constant for $M>1$. 
}
\end{figure}

We run the three considered algorithms with $N=10^6$. 
Figure \ref{fig:mixed_gaussian_posterior} shows that the MC and QMC approximations 
match closely; for this plot, $\epsilon=0.01$ (leading to a proportion of non-zero 
weights close to $10^{-3}$), and $d=1$. 

\begin{figure}[H]
\centering
\begin{subfigure}{.5\textwidth}
  \centering
  \includegraphics[width=1\linewidth]{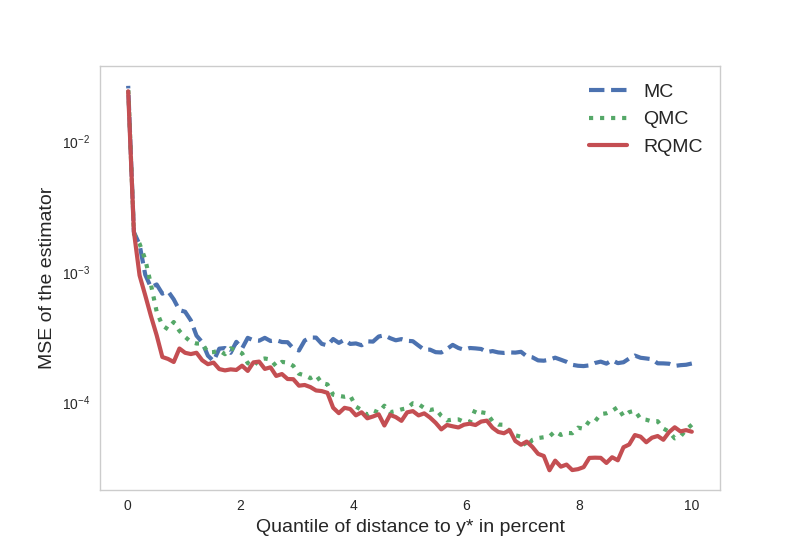}
\caption{}
  \label{fig:vars_of_means_mixed_gaussian}
\end{subfigure}%
\begin{subfigure}{.5\textwidth}
  \centering
  \includegraphics[width=1\linewidth]{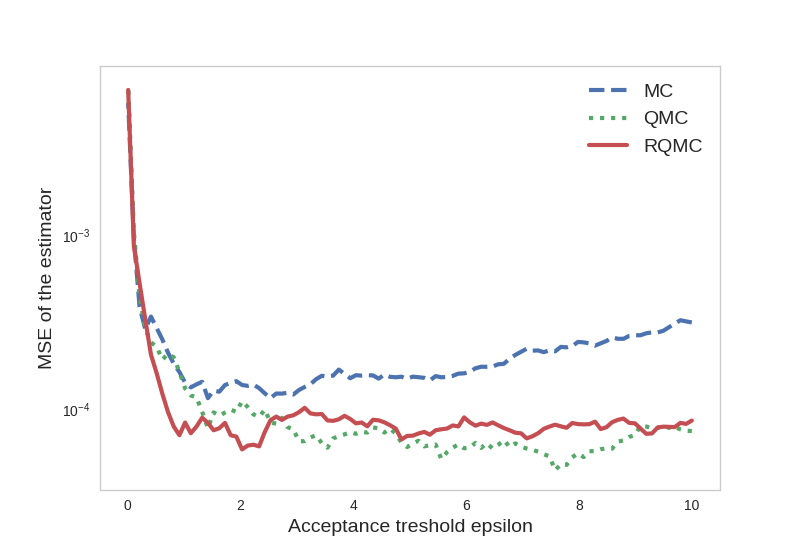}
\caption{}
  \label{fig:vars_of_vars_mixed_gaussian}
\end{subfigure}
\caption{MSE of posterior estimates as $\epsilon$ varies (Left: ABC 
	posterior mean; Right: ABC posterior variance). 
	The plots are based on $50$ runs, with $N=10^6$ simulations and $d=2$. The $x-$axis 
	corresponds to a varying $\epsilon$, which is set so that the proportion 
	of non-zero weights (i.e. the proportion of simulated $y_n$ such 
  that $\delta(y_n,y^\star)\leq \epsilon$) varies from 0 to $10\%$.
  (R)QMC sequences lead to a reduced MSE. The effect
  vanishes as $\epsilon$ goes to $0$. 
	}
\label{fig:var_mean_var_var_estimator}
\end{figure}

Figure \ref{fig:var_mean_var_var_estimator} compares the empirical variance
(over 50 runs) obtained with the three considered approaches, as a function of
$\epsilon$, when estimating the expectation (left pane) and variance (right
pane) of the ABC posterior. Here, $N=10^6$, $d=2$, and 
$\epsilon$ is chosen so as to generate a proportion of non-zero weights that 
vary from 0 to $10\%$. 

\begin{figure}[H]
\centering
\begin{subfigure}{.45\textwidth}
  \centering
  \includegraphics[width=1\linewidth]{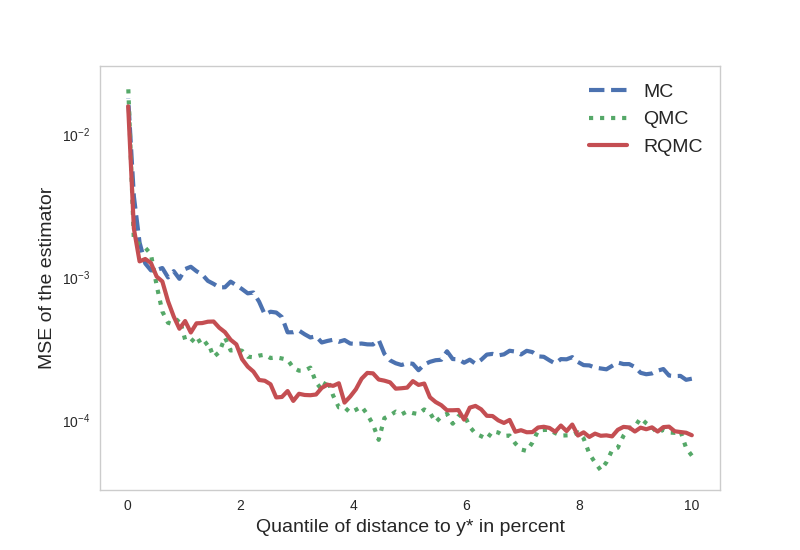}
  \caption{Left}
  \label{fig:mix_gaussian_dim4}
\end{subfigure}%
\begin{subfigure}{.45\textwidth}
  \centering
  \includegraphics[width=1\linewidth]{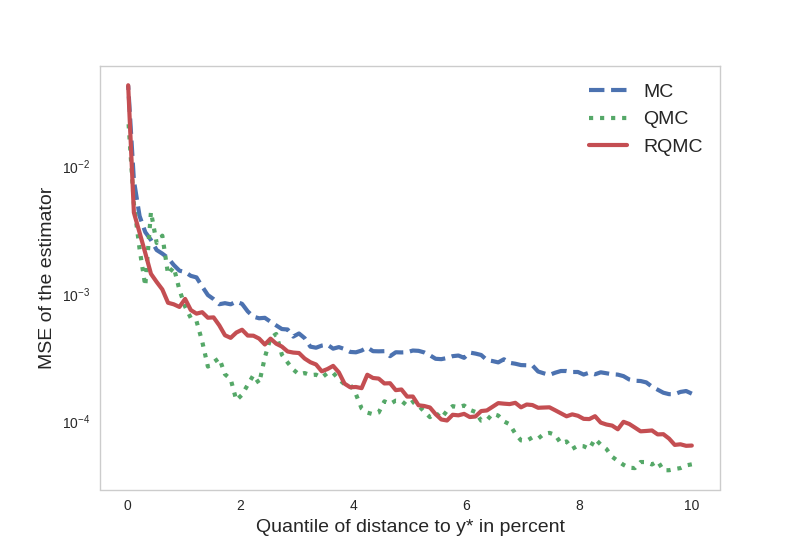}
  \caption{Right}
  \label{fig:mix_gaussian_dim8}
\end{subfigure}
\caption{Same caption as for Figure \ref{fig:vars_of_vars_mixed_gaussian}, 
	except left (resp. right) panel corresponds to $d=4$ (resp. $d=8$); 
	posterior estimate is the ABC posterior expectation in both cases. 
}
\label{fig:mix_gaussian_dim48}
\end{figure}

We observe a variance reduction when using either QMC or RQMC and 
for not too small values of $\epsilon$, but the variance reduction vanishes
as $\epsilon\rightarrow 0$. However, interestingly, the variance reduction 
(again for not too small values of $\epsilon$) remains significant when 
we increase the dimension, see Figures \ref{fig:mix_gaussian_dim48}.
(For $d>1$, the considered estimated quantity is the
expectation of the average of the $d$ components of $\thetavec$ with respect to
the ABC posterior.)

Finally, we consider increasing $M$, so as to be able to estimate the variance 
of a given ABC estimate from a single run of Algorithm \ref{algo:abc_is}, 
when using (R)QMC, as explained at the end of Section
\ref{sub:improved_normconstant}. The considered
estimate is that of the normalising constant of the ABC posterior. 
We see that the variance estimate is fairly stable even 
for small values of $M$, and that it is close to the actual variance 
(over 75 runs) of the estimate as can be seen in Figure \ref{fig:variance_during_after_several_m}.

Note that both quantities are multiplied by $M$ in Figure \ref{fig:variance_during_after_several_m}. 
This allows us to check that the adjusted variance (accounting for CPU time)
remains constant, as expected. As already explained, this means that taking 
$M>1$ is not sub-optimal (in terms of the variance vs CPU time trade-off), 
while it allows us to estimate the variance of any estimate obtained 
from the (R)QMC version of Algorithm \ref{algo:abc_is}.

\subsection{Lotka-Volterra-Model}
The Lotka-Volterra model, see \cite{Toni187}, is commonly used in population
dynamics to study the interaction in predator-prey models, for example. The
model is characterized by the respective size of the populations evolving over
time and denoted by $(X_1, X_2)$, taking values in $\mathbb{Z}^2$. 

There are three possible transitions: the prey (denoted by $X_1$) may grow by
one entity with rate $\alpha$, a predation may happen with rate $\beta$, that
reduces the prey by one unit and increases the predator population (denoted by
$X_2$) by one unit, or the predator may die with rate $\gamma$. The system is
summarized by the following rate equations: 
\begin{eqnarray*}
	(X_1, X_2) \stackrel{\alpha}{\rightarrow} (X_1+1, X_2), \\
(X_1, X_2) \stackrel{\beta}{\rightarrow}  (X_1-1, X_2+1), \\
(X_1, X_2) \stackrel{\gamma}{\rightarrow} (X_1, X_2-1), 
\end{eqnarray*}
with the corresponding hazard rates $\alpha X_1$, $\beta X_1 X_2$ and $\gamma
X_2$, respectively. The hazard rates characterize the instantaneous probability
that the system changes to a new state. The parameter of the model is 
$\thetavec=(\alpha, \beta, \gamma)$. 
The initial population is fixed to $ (50, 100)$. 

We simulate from the model using Gillespie's algorithm, see 
\cite{Toni187}, for $T=30$ time steps, and  record the size of the
population at times $t_i = 2i$, where $i = {0,\cdots,15}$. This gives two discrete
time series of length $16$. As a distance function for comparing our true
observation and the pseudo-observations, we use the Euclidean norm $\| \cdot
\|_2$ applied to the differences of the series. As a prior we use 
$\uvec \sim \mathcal{U}[-6,2]^3$, which is then 
transformed to $\thetavec = \exp(\uvec)$. 

As in the previous section, we compare the empirical variance over 50 runs 
of a given estimate obtained from the different approaches. The estimated 
quantity is the expectation of $(\alpha+\beta+\gamma)/3$ with respect to the ABC
posterior. 

\begin{figure}[H]
\centering
\begin{subfigure}{.5\textwidth}
  \centering
  \includegraphics[width=1\linewidth]{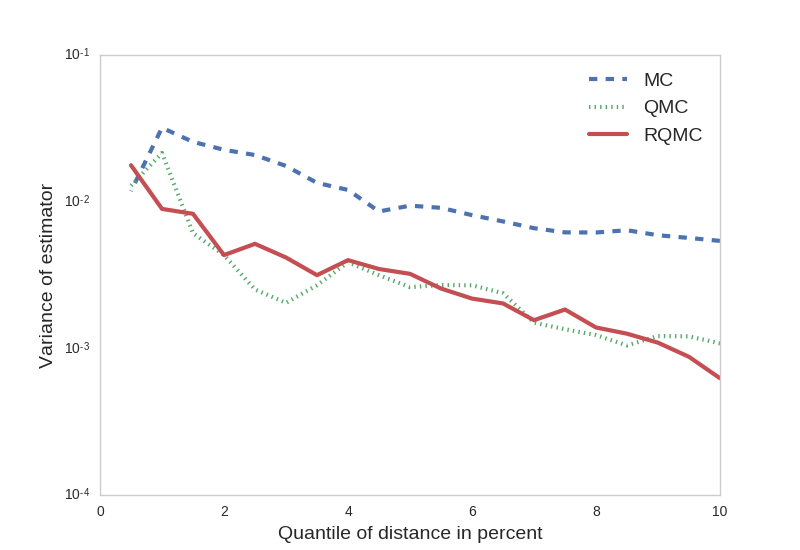}
\caption{}
  \label{fig:mean_variance_lotka-volterra}
\end{subfigure}%
\begin{subfigure}{.5\textwidth}
  \centering
  \includegraphics[width=1\linewidth]{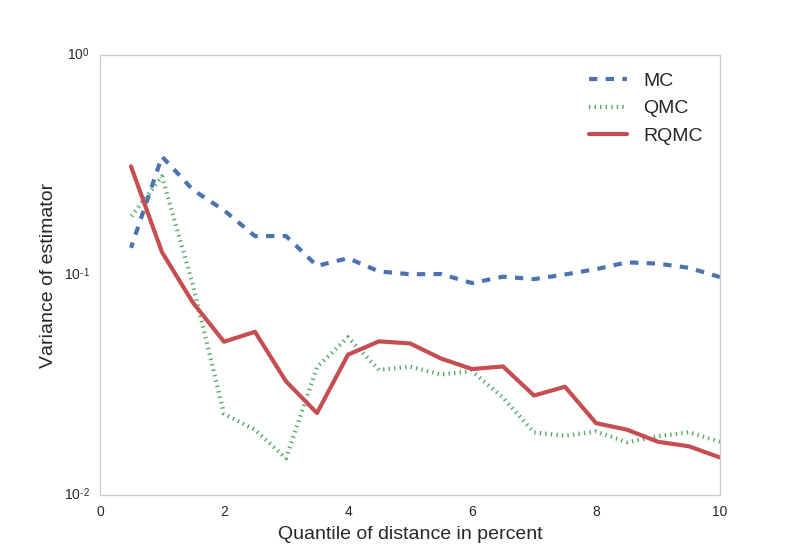}
\caption{}
  \label{fig:variance_variance_lotka-volterra}
\end{subfigure}
\caption{Variance of the mean and variance estimator for the Lotka--Volterra
model. The plots are based on $50$ repetitions of $10^5$ simulations from the prior
and the model. The accepted observations correspond to quantiles based on the
smallest distances $\delta(y_n,\ystar) $. Left: Variance of the posterior
mean estimator. Right: Variance of the posterior variance estimator} 
\label{fig:lotka_volterra}
\end{figure}

We observe the same phenomenon as in the previous example: the variance 
reduction brought by either QMC or RQMC is significant for not too small 
values of $\epsilon$, but it vanishes as $\epsilon\rightarrow 0$. 

\subsection{Tuberculosis mutation} \label{ssec:tuberculosis_mutation}

The following application is based on the estimation of tuberculosis
reproduction rates as in \cite{tanaka2006using}. The interest lies in
recovering the posterior distribution of birth, death and mutation rates
$(\alpha, \beta, \gamma)$ of a tuberculosis population that has been recorded
in San Francisco over a period from 1991 to 1992. 

The simulator of the model is based on an underlying continuous time Markov
process where $t$ denotes the time and $N(t)$ denotes the size of the
population. Starting from one single bacterium the individual can either
replicate itself with rate $\alpha$, die with rate $\gamma$ or mutate to a new
genotype with rate $\beta$. The number of bacteria having the same genotype is
recorded at every step and the simulation is run forward until a size of $N(t)
= 10^4$ has been obtained. At every step in the simulation a bacterium is
chosen uniformly at random and one of the three events $(\alpha, \beta,
\gamma)$ is applied to it. After simulating a population of $10^4$ bacteria,
the simulation is stopped and a subpopulation of $473$ bacteria is sampled. The
ensuing population is characterized by the cluster size of bacteria that have
the same genotype. The data is available in Table \ref{table:tuberculosis}. For
instance, there were $282$ clusters with only one bacterium with the same
genotype and there were $20$ clusters that contained two bacteria with the same
genotype.

\begin{table}[H]
\centering
\begin{tabular}{ccccccccccc}
Cluster size & 1 & 2 & 3 & 4 & 5 & 8 & 10 & 15 & 23 & 30 \\ 
Number of clusters & 282 & 20 & 13 & 4 & 2 & 1 & 1 & 1 & 1 & 1 \\ 
\hline 
\hline 
\end{tabular} 
\caption{Tuberculosis bacteria genotype data}
\label{table:tuberculosis}
\end{table}

\begin{figure}[H]
\centering
\begin{subfigure}{.5\textwidth}
  \centering
  \includegraphics[width=0.9\linewidth]{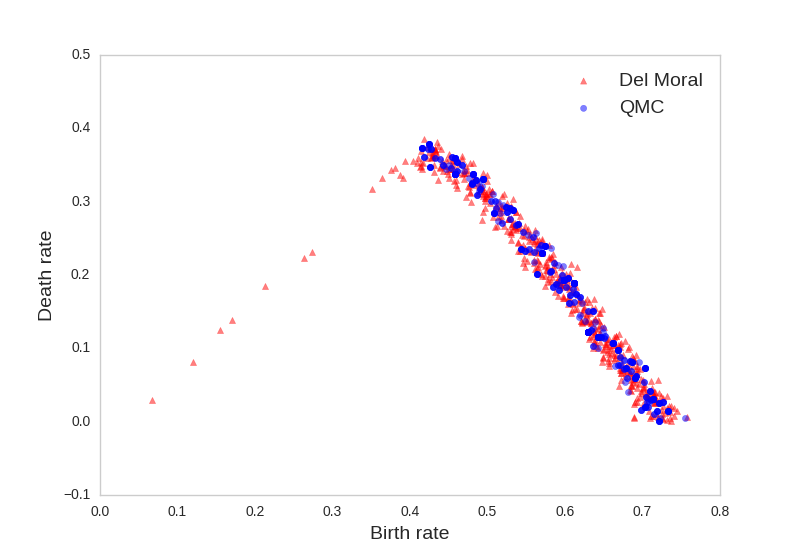}
\caption{}
  \label{fig:tuberculosis2}
\end{subfigure}%
\begin{subfigure}{.5\textwidth}
  \centering
	\includegraphics[width=0.9\linewidth]{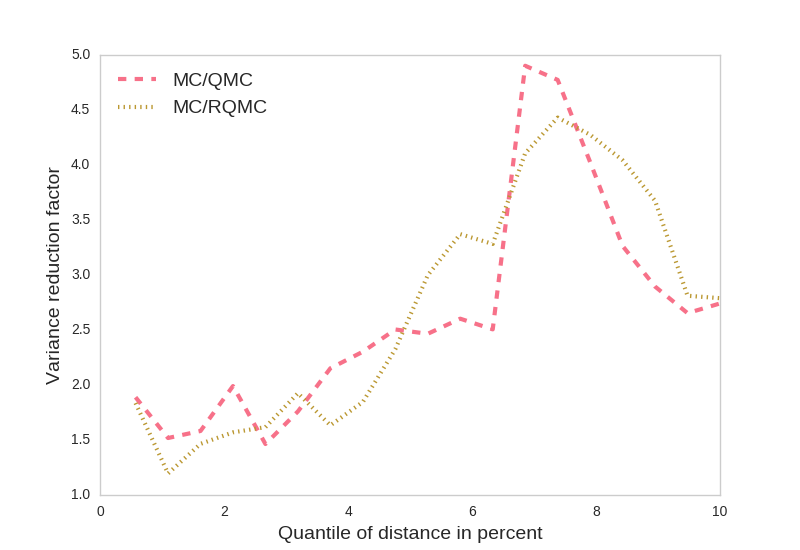}
\caption{}
	\label{fig:tuberculosis_static}
\end{subfigure}
\caption{
Left: Posterior distribution of the tuberculosis mutation
model. The x--axis corresponds to birth rate $\alpha$, the y--axis corresponds
to the death rate $\beta$, $N=500$. Right: Variance reduction factors (computed from 
50 runs based on $N=10^4$) as a function of the proportion of non-zero weights.} 
\label{fig:tuberculosis_general}
\end{figure}

The parameters must satisfy the conditions $\alpha+\beta+\gamma=1$, $0\leq
\alpha,\beta, \gamma \leq 1$, and $\alpha > \gamma$. (The last constraint 
prevents the population from dying out.) Thus, we let $\beta=1-\alpha-\gamma$, 
and assign a uniform prior to $(\alpha, \gamma)$, subject to $\alpha>\gamma$. 
 \cite{tanaka2006using}
used as a summary statistic for the data the quantities $y= (g/473, 1-\sum_i
(n_i/473)^2)$, where $g$ denotes the number of distinct clusters in the sample
and $n_i$ is the number of observed bacteria in the $i$th genotype cluster. The
distance between a pseudo observation and the observed data is finally
calculated as the Euclidean distance between $y$ and $\ystar$. 
Figure \ref{fig:tuberculosis2} shows the recovered posterior distribution after
application of a sequential sampling approach, that is described in Section
\ref{sec:sequential_applications}. We see our method, denoted by QMC and the
method of \cite{DelMoral2012}, denoted by \textit{Del Moral} recover the same
posterior distribution. There remain some artifacts in the second method, due
to a slightly higher acceptance threshold $\epsilon=0.12$ compared to
$\epsilon=0.08$ as in the QMC approach.
We estimate the ABC posterior expectation of $(\alpha+\gamma)/2$ 
and then compare the empirical variance of this
estimator. The result of the repeated simulation of this estimator is shown in
Figure \ref{fig:tuberculosis_static}, where we show the value of
$\Var_{MC}/\Var_{(R)QMC}$, where $\Var_{MC}$ is the variance of the posterior
estimator based on a MC sequence. This quantity allows to assess the variance
reduction factor as a function of the acceptance threshold. Again, we observe a
declining variance reduction as $\epsilon\rightarrow 0$. 
Nevertheless, the variance reduction even for the smallest acceptance threshold
is still of factor $1.5$, which means that we need $33\%$ fewer simulations in
order to achieve the same precision of the estimator.

\subsection{Concluding remarks}

As predicted by the theory, we observed that using QMC (or RQMC) to generate 
the parameter values (in Algorithm \ref{algo:abc_is})
always reduce the variance of ABC estimates. 
However, the variance reduction becomes small when $\epsilon \rightarrow 0$. 
But it should be noted that any static ABC algorithm, such as Algorithm
\ref{algo:abc_is} becomes very wasteful when $\epsilon$ is small, 
as most simulated datapoints lies outside 
the ball defined by the constraint $\delta(y, y^\star) \leq \epsilon$ in such 
a case. 
In order to take $\epsilon$ smaller and smaller, it seems to make more sense 
to progressively refine the proposal distribution, based on past simulations. 
This is the point of sequential ABC algorithms, which we discuss in the next
two sections.

\section{Sequential ABC} \label{sec:sequential_sampling}

\subsection{Adaptive importance sampling}
One major drawback of Algorithm \ref{algo:abc_is} is that the quality of the
approximation in \eqref{eq:is_estimator} depends on how well the proposal
distribution $q(\thetavec)$ matches the target distribution
$p_\epsilon(\thetavec) $. If, for example, the proposal is very flat and the
target is spiky due to a small value of $\epsilon$, only a small number of
particles will cover the region of interest. The idea of sequential
ABC algorithms is therefore to sequentially decrease $\epsilon$ over a range
of time steps $t \in 0:T$ while adapting the proposal  distribution
$q_t(\thetavec)$ so as to make it closer and closer to the true posterior. 

In the current setting we will use a flexible parametric approximation
$q_{t}(\thetavec)$ of the ABC posterior $p_{\epsilon_{t}}(\thetavec)$, that is
estimated from the the samples $(\thetavec^{t-1}_n, w^{t-1}_n)_{n \in
1:N}$. This distribution $q_{t}(\thetavec)$ is then used to simulate new
particles $(\thetavec^{t}_n)_{n \in 1:N}$. 
The corresponding algorithm is given as pseudo-code in Algorithm
\ref{algo:abc_amis}.

\begin{algorithm}[H] \label{algo:abc_amis}
\SetKwInOut{Input}{Input}
 \Input{Observed $\ystar$, prior distribution $p(\thetavec)$, 
 simulator $q_{\thetavec} (x)$, 
 initial threshold $\epsilon_0$, 
 number of simulations $N$, weighting procedure $\pest(x)$.}
 \KwResult{Set of weighted samples $(\thetavec^t_n, x^t_n, w^t_n)_{n \in 1:N, t \in 0:T}$}
 \For{$n = 1$ to $N$}{

	 Sample $\theta_n^0\sim p(\thetavec)$ \;
	 set $w_n^0 = 1$ \;
 }
 \For{$t = 1$ to  $T$ }{
	 Set $\epsilon_{t}$ and  $q_{t}(\thetavec)$ based on 
	 $(\thetavec^{t-1}_n, x^{t-1}_n, w^{t-1}_n)_{n \in 1:N}$ \;
 \For{$n=1$ to $N$}{
  Sample $\thetavec^{t}_n \sim q_{t}(\thetavec)$ \;
  Sample $x^{t}_n \sim q_{\thetavec_n^{t}}(x)$ \;
  Set $w^{t}_n = p(\thetavec^{t}_n) \pestt(x^{t}_n)/q_t(\thetavec^{t}_n)$ \;
 }
 }
 \caption[ABC_accept_reject]{ABC adaptive importance sampling algorithm}
\end{algorithm}

\subsection{Adapting the proposal $q_t$}

\subsubsection{Gaussian proposal}\label{sub:single_Gaussian}

The simplest strategy one may think of to adapt $q_t$ is to set it to 
a Gaussian fit of the previous weighted sample.
Although basic, we shall see that this approach tends to work well in 
practice, unless of course the actual posterior is severely multimodal, strongly skewed
or has heavy tails. 

\subsubsection{Mixture of $N$ components}

The sequential Monte Carlo sampler (SMC) of 
\cite{Sisson2009} may be viewed as a particular version of Algorithm 
\ref{algo:abc_amis}, where $q_t$ is set to a mixture of $N$ Gaussian components
centred on the $N$ previous particles $\thetavec^{t-1}_n$, with 
covariance matrix $\hat{\Sigma}^{t-1}$ set to
twice the empirical covariance of these particles. The proposal distribution reads
\begin{eqnarray*}
	q_{t}(\thetavec) = \frac{\sum_{n=1}^N w_n^{t-1}
	\mathcal{N}(\thetavec | \thetavec^{t-1}_n,2\hat{\Sigma}^{t-1})}{\sum_{n=1}^N w_n^{t-1}}.
\end{eqnarray*}

This results in an algorithm
of complexity $\mathcal{O}(N^2)$ since for every proposed new particle
$\thetavec^{t}_n$, computing the corresponding weight involves a sum 
over $N$ terms. 

\subsubsection{Mixture proposal with a small number of components}
As an intermediate solution between a single Gaussian distribution and a
mixture of $N$ Gaussian distributions, 
we suggest to use a Gaussian mixture with a small number of components. 
We suggest to estimate the mixture  via a Variational Bayesian
procedure, see \cite{blei2016variational}, but other methods as Expectation
Maximization could also be used. The
proposal distribution reads 
\begin{eqnarray*}
q_{t}(\thetavec) =  \sum_{j=1}^J \alpha_j^{t-1} \mathcal{N}(\thetavec |
\hat{\mu}_j^{t-1}, \lambda \hat{\Sigma}_j^{t-1}), 
\end{eqnarray*}
where $\alpha_j^{t-1}$, 
$\hat{\mu}_j^{t-1}$, and 
$\hat{\Sigma}_j^{t-1}$
denote respectively 
the weight, mean, and covariance matrix of cluster $j$ estimated at iteration 
$t-1$. Again, we artificially inflate the covariances with a factor $\lambda >
1$ in order to put more mass in the tails of the proposal distribution. 
In our numerical experiments we set $\lambda=1.2$. 
Regarding $J$, we may either fix it arbitrarily or use the Variational Bayesian 
approach to choose it automatically. 

In order to generate QMC or RQMC points from such a mixture distribution, 
we set the number of samples for each cluster $j$ to $N_j^t = \left
\lfloor{\alpha_j^{t-1} N}\right \rfloor $ and potentially adjust $N_j^t$ as to make
sure that $\sum_j N_j^t = N$ holds. For each cluster $j$, 
a (R)QMC sequence of length $N_j^t$ is
generated and transformed to the sample of a Gaussian distribution
$\mathcal{N}(\thetavec | \hat{\mu}_j^{t-1}, \lambda \hat{\Sigma}_j^{t-1})$. This is
achieved via the transformation of the (R)QMC sequence $(\uvec_n)_{n \in
1:N_j^t}$ via the component-wise quantile function $\Phi^{-1}(\cdot)$:
$\thetavec^t_n = \hat{\mu}_j^{t-1}
+ C_{t-1}\Phi^{-1}(\uvec_n)$, where $C_{t-1}$ is the Cholesky triangle of the covariance 
matrix: $C_{t-1} (C_{t-1})^T = \lambda \hat{\Sigma}_j^{t-1}$. 

This approach has the following advantages. First, we maintain flexibility by
allowing to cover several modes, as the posterior distribution might be
multi--modal. Second, the use of a limited number of clusters makes sure that
we can benefit from the better coverage of the space that comes from the use of
(R)QMC sequences. Using only a small number of clusters preserves the structure
of the (R)QMC point set. Other approaches based on the inverse Rosenblatt transform
\citep{gerber2015sequential} are computationally more expensive. 
In contrast, using the approach of \cite{Sisson2009}
would destroy
the properties of the low discrepancy or scrambled net sequences and hence the
variance reduction that comes from the (R)QMC sequence could vanish. (This has
been found as a result of our simulation studies, not shown here.)

\subsection{Adapting simultaneously $\epsilon_t$ and the number of simulations 
per parameter}\label{sub:hybrid}

As discussed in Section \ref{sub:pseudo}, the weights $\pestt(x^{t}_n)$ are unbiased 
estimators of the probabilities $P_{\thetavec_n^t}(\delta(y^\star,y)\leq \epsilon_t)$, 
which may be obtained in two ways: (a) as an average over a fixed number $M$ 
of simulations; or (b) as a function of the number of simulations required
so that $k$ of them are at a $\epsilon$ distance of $y^\star$; that random 
number follows a negative binomial distribution. 

So far, we have focused on (a), and even took $M=1$ in our first set of 
numerical examples in Section \ref{sec:non_sequential_applications}. 
If we use this strategy, we may follow \cite{DelMoral2012} in 
adapting $\epsilon_t$ according to the ESS \citep[effective sample size,][]{KongLiuWong}; 
i.e. at iteration $t$, once we have simulated the $\thetavec_n^t$'s and the $x_n^t$'s, 
we solve numerically (using bisection) in $\epsilon_t$
the equation $\mathrm{ESS}=\alpha N$, for $\alpha\in(0,1)$, where 
\[ \mathrm{ESS} =  \frac{(\sum_{n=1}^N w_n^t)^2}{\sum_{n=1}^N (w_n^t)^2}
\]
and $w_t^n = p(\theta_n^t) \hat{L}_\epsilon(x_n^t)/q(\theta_n^t)$, 
$\hat{L}_\epsilon(x_n^t) = M^{-1} \sum_{m=1}^M 
\ind \{\delta(y^\star,y_{n,m}^t) \leq \epsilon_t \}$.

This approach usually works well during the first iterations 
of Algorithm \ref{algo:abc_amis}, but it is bound to collapse as $\epsilon$
gets too small: as $\epsilon\rightarrow 0$, 
$P_{\thetavec}(\delta(y^\star,y)\leq\epsilon)\rightarrow 0$
whatever $\thetavec$, and as a result most weights $w_t^n$ become zero when
$\epsilon_t$ is too small. One remedy is to set $M$ to a much larger value, 
so that weights take much longer to collapse. However, this is expensive 
and wasteful, given that the first iterations would work well with a much 
smaller $M$. 

In that sequential context, the negative binomial strategy for computing the 
weights becomes appealing, as it makes it possible to adapt automatically the CPU 
effort to a given $\epsilon$: we may decrease $\epsilon_t$ at each iteration, 
while ensuring that the variance of the weights (as estimates of the 
probabilities $P_{\thetavec_n^t}(\delta(y^\star, y)\leq\epsilon_t)$ does not blow up.
Of course, the price to pay is that iterations become more and more expensive.

In practice, we found that that this approach was unwieldy during the first 
iterations of the algorithm: during that time, a few 
simulated parameters $\thetavec_n^t$ are such that the corresponding probability 
that $\delta(y^\star, y)\leq \epsilon_t$ is much smaller than for the other 
particles. As a result, the negative binomial estimate requires generating 
a lot of observations for those particles, which typically gets discarded 
later.

Thus, in the end, we recommend the following hybrid strategy:

\begin{itemize}
	\item At iterations $t=0$ to $t=T_1$ (say $T_1=10$), use the `fixed
		$M$' (say $M=10$) strategy to compute the weights, and 
		adapt $\epsilon_t$ using the ESS. 
	\item At iterations $t>T_1$, switch to the negative binomial strategy
		for computing the weights, and adapt $\epsilon_t$ as follows: 
		set it to the median of the distance values $\delta(y^\star,y_{n})$ 
		where the $y_n$'s denote here all the artificial observations 
		generating during the previous iteration such that 
		$\delta(y^\star, y_n)\leq \epsilon_{t-1}$. Stop when 
		$\epsilon_t$ gets below a certain target value $\epsilon^\star$. 
\end{itemize}

\section{Numerical illustration of the sequential procedure}
\label{sec:sequential_applications}

\subsection{Toy model}
We return to the toy model of Section \ref{ssec:toy_model_non_sequential}, 
taking this time $d=3$. We compare five algorithms: 
three versions of Algorithm \ref{algo:abc_amis} with the $\thetavec_n^t$'s 
generated using, respectively, Monte Carlo, Quasi-Monte Carlo, and 
RQMC; the sequential ABC algorithm of \cite{Sisson2009}, which 
(as explained previously) is essentially Algorithm \ref{algo:abc_amis} 
with a mixture proposal with $N$ components; 
and finally the algorithm of \cite{DelMoral2012}. (The algorithm of
\cite{DelMoral2012} generates the $\thetavec_n^t$ by evolving the 
particles resampled at the previous iteration through a Markov kernel; 
see the paper for more details.) 

Regarding the adaptive choice of $\epsilon_t$, 
we use the hybrid strategy outlined in the previous section 
for our MC, QMC and RQMC algorithms, we use the ESS-based strategy for 
\cite{DelMoral2012}'s algorithm, and we use the following strategy for
\cite{Sisson2009}'s: $\epsilon_t$ is set to the median of the distances 
$\delta(y^\star,y_n)$ computed at the previous iteration. For all these 
algorithms, we set $M=10$.

\begin{figure}[H]
\centering
\begin{subfigure}{.5\textwidth}
  \centering
  \includegraphics[width=1\linewidth]{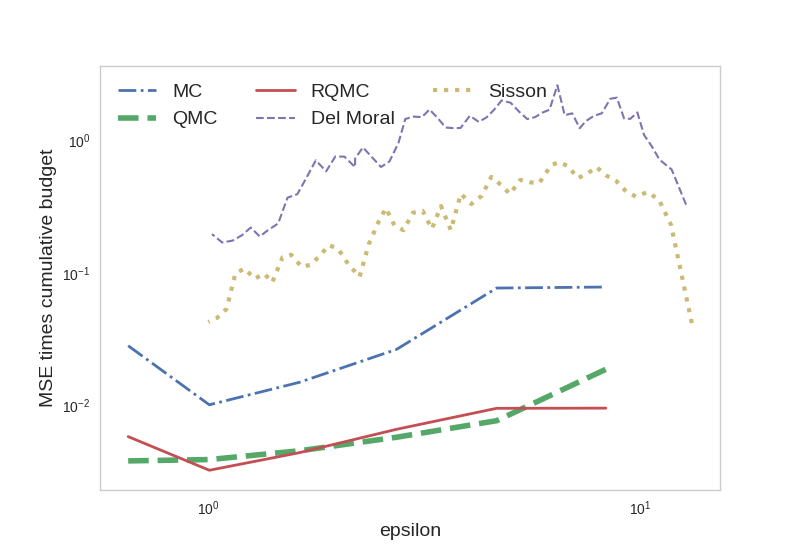}
\caption{}
  \label{fig:sequential_vars_of_means_mixed_gaussian}
\end{subfigure}%
\begin{subfigure}{.5\textwidth}
  \centering
  \includegraphics[width=1\linewidth]{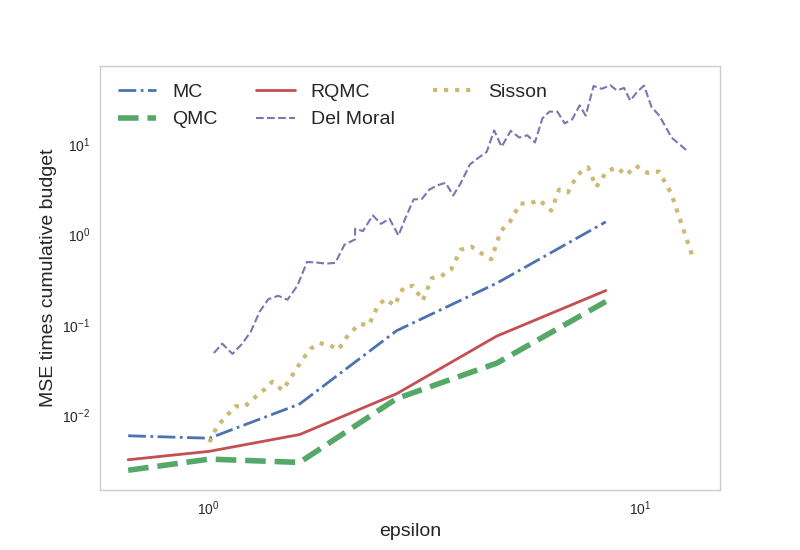}
\caption{}
  \label{fig:sequential_vars_of_vars_mixed_gaussian}
\end{subfigure}
\caption{three-dimensional Gaussian toy example. Algorithms run with $N=10^3$ particles.
	Adjusted MSE (as defined in the text) at iteration $t$, as function 
	of $\epsilon_t$, for the following posterior estimate: 
	exceptation (left) and variance (right) of
	$\bar{\thetavec}=(\theta_1+\theta_2+\theta_3)/3$.}
\label{fig:sequential_mixed_gaussian_1small}
\end{figure}

\begin{figure}[H]
\centering
\begin{subfigure}{.5\textwidth}
  \centering
  \includegraphics[width=1\linewidth]{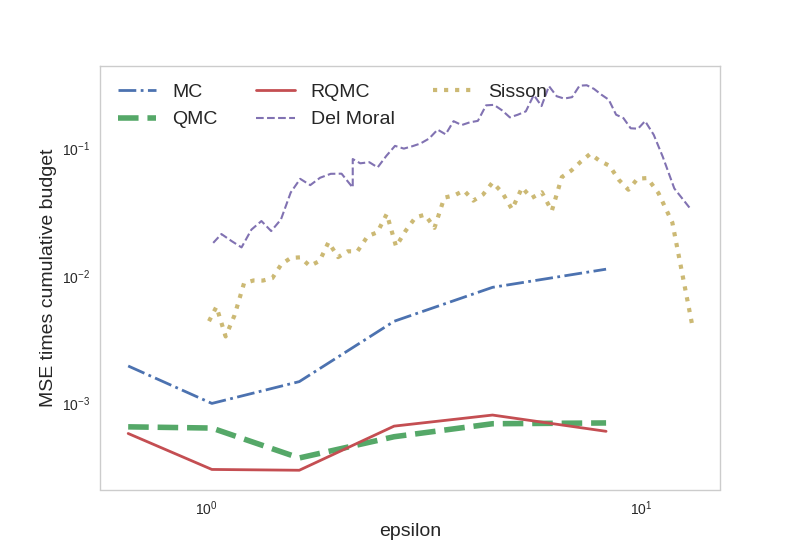}
\caption{}
  \label{fig:sequential_vars_of_means_mixed_gaussian}
\end{subfigure}%
\begin{subfigure}{.5\textwidth}
  \centering
  \includegraphics[width=1\linewidth]{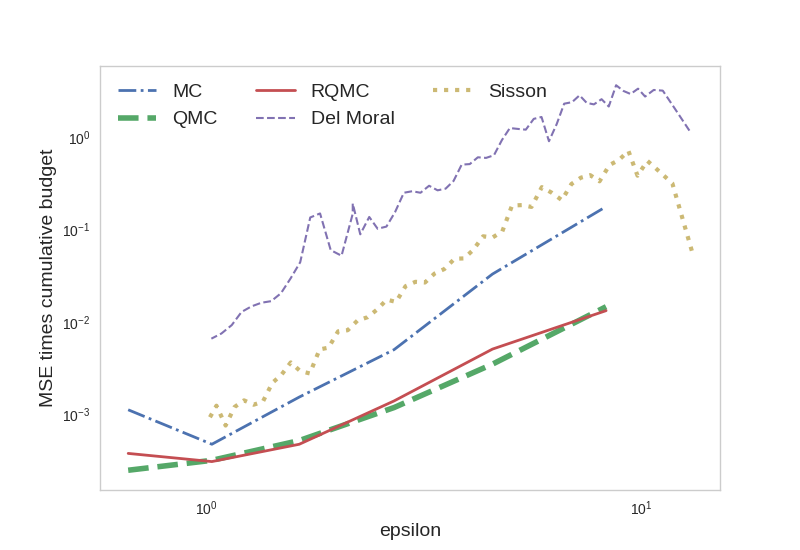}
\caption{}
  \label{fig:sequential_vars_of_vars_mixed_gaussian}
\end{subfigure}
\caption{Same as Figure \ref{fig:sequential_mixed_gaussian_1small}, except
algorithms are run with $N=10^4$ particles.}
\label{fig:sequential_mixed_gaussian_1large}
\end{figure}

For this toy model, we simply consider the basic strategy for adapting $q_t$
outlined in Section \ref{sub:single_Gaussian}, i.e. $q_t$ is a Gaussian fit to the 
previous set of particles. The five algorithms are run with either $N=10^3$ 
(Figure \ref{fig:sequential_mixed_gaussian_1small}) or 
$N=10^4$ particles (Figure \ref{fig:sequential_mixed_gaussian_1large}); in both 
cases the algorithms are stopped when $\epsilon_t\leq \epsilon^\star = 1$.
In both figures, we plot the adjusted MSE at iteration $t$ as a function 
of $\epsilon_t$, where the adjusted MSE is the empirical MSE
of a given estimate (over 50 runs) times the number of observations generated 
from the model up to time $t$. The adjusted MSE make it possible to 
account for the different running times of the algorithms. 
See also Table \ref{table:seq_toy} for a direct comparison in terms of 
both CPU effort and MSE. 

The considered estimates are the same as in Section 
\ref{ssec:toy_model_non_sequential}, 
i.e. the ABC posterior expectation and variance of $\bar{\thetavec}$, the 
average of the components of vector $\thetavec$. At least for posterior 
expectations, we see that the QMC and RQMC versions outperform the MC 
version of our algorithm, which in turn outperforms the sequential ABC algorithms 
of \cite{Sisson2009} and \cite{DelMoral2012}. 

\begin{table}[H]
\centering
\begin{tabular}{cccccc}
Sampling method	& MSE $\overline{\theta}$ & MSE
$\overline{\Var{\theta}}$ &  number simulated datapoints & 
$\epsilon_T$  \\ 
\hline 
AIS-MC 	& 0.00162 & 0.00037 & 44,980 & \textbf{0.65}\\ 
AIS-QMC & \textbf{0.00039} & 0.00014 & \textbf{32,919} & \textbf{0.65}\\ 
AIS-RQMC & 0.00049 & 0.00013  & 42,088 & \textbf{0.65}\\ 
Del Moral & 0.00117 & 0.00018 & 580,000 & 1.0 \\  
Sisson & 0.00117 & \textbf{0.00010} & 125,928 & 0.95\\ \hline 
IS-MC & 0.00128 & 0.00513 & 1,000,000 & 0.65\\ \hline 
\hline 
\end{tabular} 
\caption{Toy example, performance of the five considered 
sequential algorithms at the final iteration $T$, for $N=10^3$ particles. 
IS-MC corresponds to 
the plain IS sampling without adaptation.}
\label{table:seq_toy}
\end{table}

\subsection{Bimodal Gaussian distribution}

In order to illustrate the flexibility that comes from using a mixture of
Gaussians for the proposal we consider a model that yields a
multi-modal posterior: 
\begin{eqnarray*}
\thetavec &\sim& \mathcal{U}\left([-10,10]^d \right) , \\
y_i &\stackrel{iid}{\sim}& \frac{1}{2}\mathcal{N}(\thetavec, I_d) 
+ \frac{1}{2}\mathcal{N}(-\thetavec, I_d),\quad i=1,\ldots, 100.
\end{eqnarray*}
We simulate $y^\star$ from the model. Throughout this application we set $d=2$. 
The model is not identifiable and thus generates a bimodal posterior. 
Regarding the distance $\delta$, we follow the idea
of \cite{Bernton2016} and use the optimal transport distance between $\yvec$
and $\yvec^*$, more specifically the earth-movers-distance.

\begin{figure}[H]
\centering
\begin{subfigure}{.5\textwidth}
  \centering
  \includegraphics[width=1\linewidth]{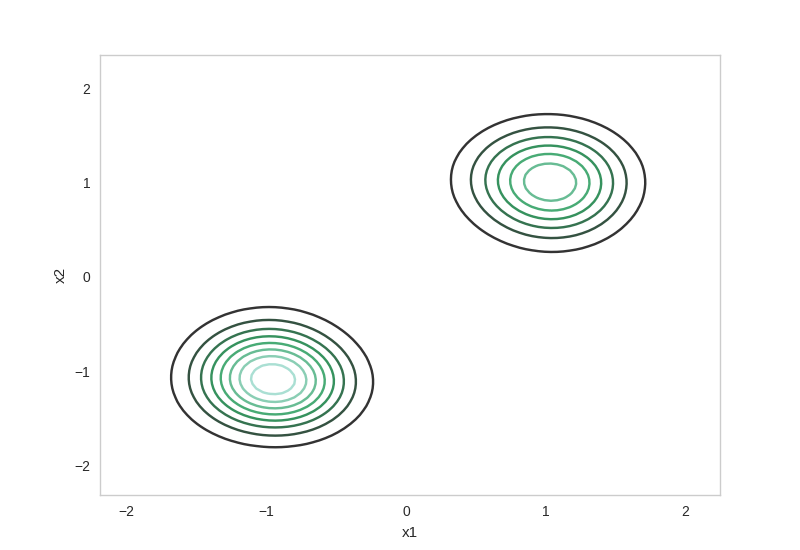}
\caption{}
  \label{fig:sequential_bimodal_posterior}
\end{subfigure}%
\begin{subfigure}{.5\textwidth}
  \centering
  \includegraphics[width=1\linewidth]{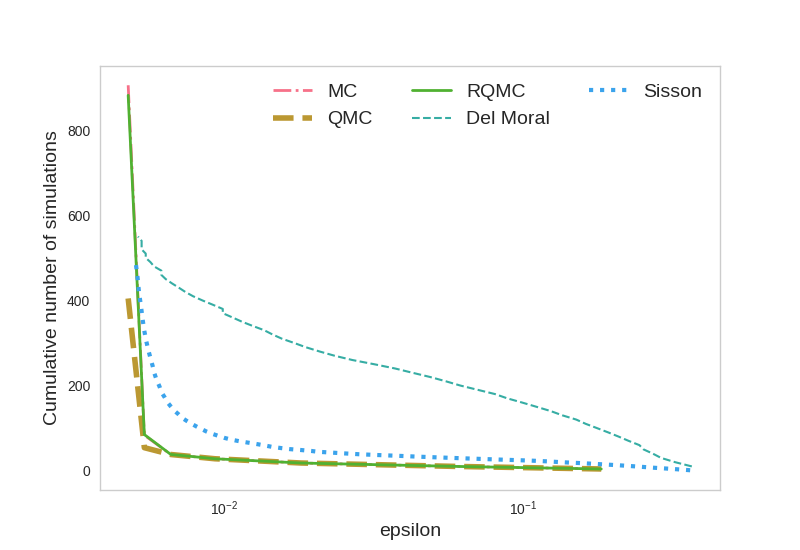}
\caption{}
  \label{fig:sequential_bimodal_cumulative}
\end{subfigure}
\caption{Simulation for the bimodal distribution. Left: recovered posterior
	distribution. Right: average (over 50 runs) 
	of cumulative number of simulations from the simulator across particles according to
acceptance threshold; algorithms were run with $N=10^3$ particles.}
\end{figure}

\begin{figure}[H]
\centering
\begin{subfigure}{.5\textwidth}
  \centering
  \includegraphics[width=1\linewidth]{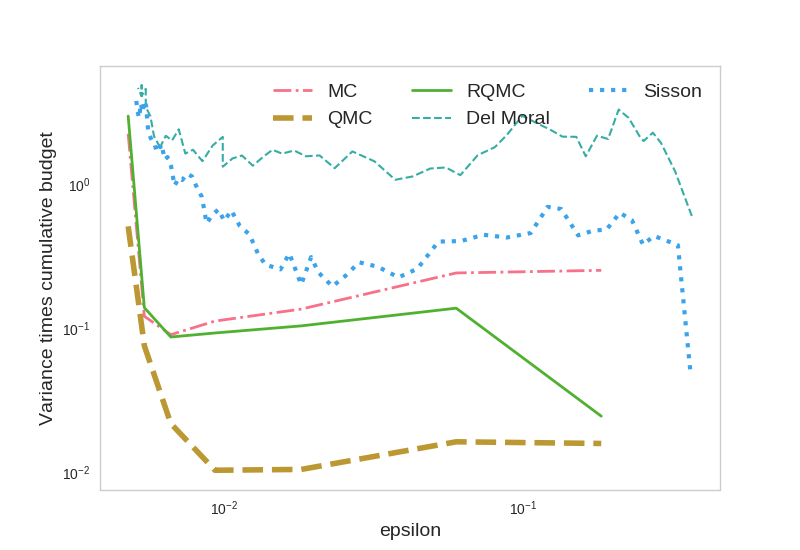}
\caption{}
  \label{fig:sequential_bimodal_mean}
\end{subfigure}%
\begin{subfigure}{.5\textwidth}
  \centering
  \includegraphics[width=1\linewidth]{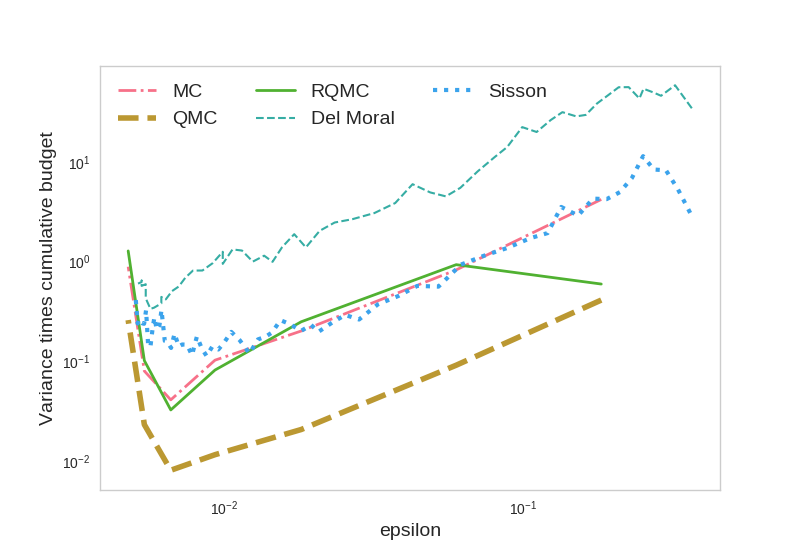}
\caption{}
  \label{fig:sequential_bimodal_var}
\end{subfigure}
\caption{Same plot as in Figure \ref{fig:sequential_mixed_gaussian_1small} 
for the bimodal example and $N=10^3$.}
\label{fig:sequential_bimodal}
\end{figure}

We set $\epsilon^\star = 5\times 10^{-3}$. 
This value has been chosen as before as a small quantile of the realized distances
after $10^6$ simulations from the prior and the simulator. The recovered
posterior is shown in Figure \ref{fig:sequential_bimodal_posterior}. Figure
\ref{fig:sequential_bimodal_cumulative} illustrates the adaptivity in the
simulation from the simulator achieved via the negative binomial approach. As
the threshold becomes smaller and smaller, the number of necessary simulations
start to increase severely. In the end, the number of necessary simulations of
the different methods catch up with each other. Still, the approaches based on
(R)QMC achieve a lower variance of the estimator as is illustrated in Figures
\ref{fig:sequential_bimodal_mean} and \ref{fig:sequential_bimodal_var}.

\subsection{Tuberculosis mutation}

\begin{figure}[H]
\centering
\begin{subfigure}{.5\textwidth}
  \centering
  \includegraphics[width=1\linewidth]{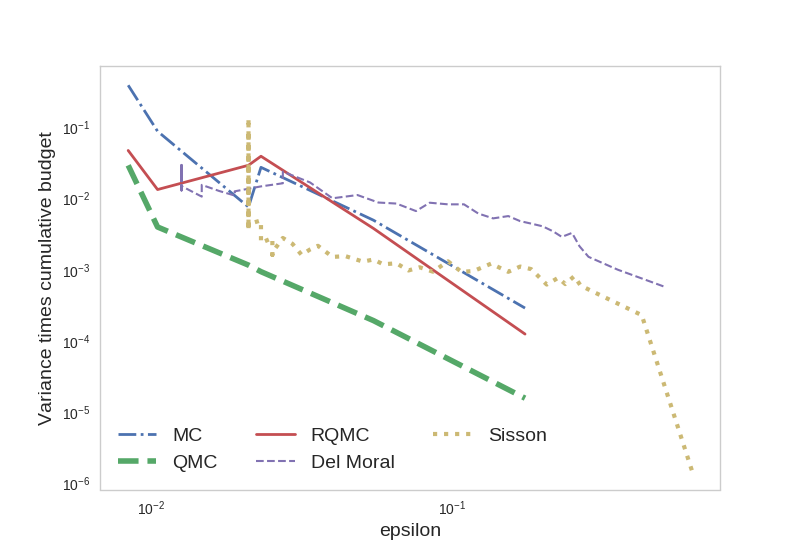}
\caption{}
  \label{fig:sequential_tuberculosis_mean}
\end{subfigure}%
\begin{subfigure}{.5\textwidth}
  \centering
  \includegraphics[width=1\linewidth]{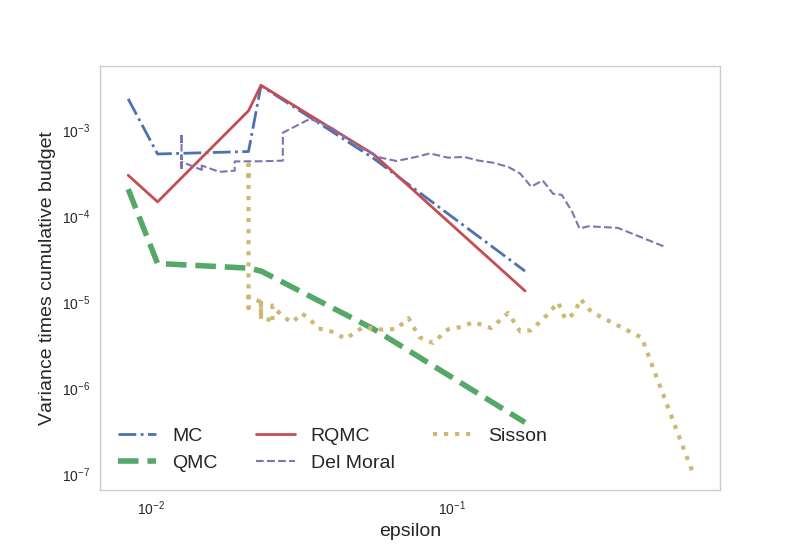}
	\caption{}
  \label{fig:sequential_tuberculosis_var}
\end{subfigure}
\caption{Same plot as in Figure \ref{fig:sequential_mixed_gaussian_1small}
for the tuberculosis example and $N=500$ particles}
\label{fig:sequential_tuberculosis}
\end{figure}

We now return to the tuberculosis example presented in Section
\ref{ssec:tuberculosis_mutation}; we set the target value $\epsilon^\star=0.01$,
and restrict the CPU budget to 
$10^6$ simulations from the model, as these simulations are
computationally intensive. We see that again the QMC approach performs best in
terms of number of simulations needed and also in terms of variance times
computational budget; see Figures \ref{fig:sequential_tuberculosis_mean}
and \ref{fig:sequential_tuberculosis_var}, and Table \ref{table:means_tuberculosis}. 
The approach of \cite{Sisson2009}
exceeds the total computation budget and thus does not reach the fixed threshold.
Figures  \ref{fig:sequential_tuberculosis_mean} and
\ref{fig:sequential_tuberculosis_var} illustrate the effect of the hybrid 
strategy for adapting $\epsilon$ and the number of simulations per parameter 
value (Section \ref{sub:hybrid}). 
The kink in the lines for the adaptive importance sampling approaches
corresponds to the moment when the weighting is obtained via the negative
binomial distribution. 

\begin{table}[H]
\centering
\begin{tabular}{cccccc}
Sampling method	& Variance $\overline{\theta}$ & Variance
$\overline{\Var{\theta}}$ & number sim. datapoints & 
$\epsilon_T$  \\ 
\hline 
AIS-MC 	& 0.376 & $5.916 \times 10^{-6}$ & 419,353& 0.008 \\ 
AIS-QMC &  0.380 & $1.156 \times 10^{-6}$ & \textbf{212,183}& \textbf{0.008} \\ 
AIS-RQMC & 0.378 & $1.001 \times 10^{-6}$ & 318,196& 0.008 \\ 
Del Moral & \textbf{0.375} & $1.065 \times 10^{-6}$ & 495,000 & 0.010  \\  
Sisson &  0.393 & $\mathbf{1.834 \times 10^{-7}}$& 1,367,949 & 0.021 \\ \hline 
\hline 
\end{tabular} 
\caption{Tuberculosis example, performance of the five considered 
	sequential algorithms at the final iteration $T$, for $N=500$ particles}
\label{table:means_tuberculosis}
\end{table}

\section{Conclusion} \label{sec:discussion}

In this paper we introduced the use of low discrepancy sequences in approximate
Bayesian computation. We found that from both a theoretical and practical
perspective the use of (R)QMC in ABC can yield substantial variance reduction.
However, care must be taken when using (R)QMC sequences. First, the
transformation of uniform sequences to the distribution of interest must
preserve the low discrepancy properties of the point set.  This is of major
importance for a sequential version of the ABC algorithm that is based on
adaptive importance sampling.  Second, the advantage of using (R)QMC tends to
diminish with the dimension (of the parameter space).  Fortunately, the
dimension of the parameter space is often small in ABC applications.  From a
practical perspective we recommend to use RQMC point sets instead of QMC as
these allow the assessment of the error via repeated simulation.

Another contribution of this paper is the use of the negative binomial
distribution in order to adapt the CPU cost of sampling from the likelihood
(for a given $\thetavec$) to the considered threshold $\epsilon$. 
This approach seems to reduce significantly the overall CPU cost. 

Finally, If the user suspects a multimodal posterior, we recommend to estimate
a mixture distribution based on the accepted samples and generate RQMC samples
based on the mixture.

\section*{Acknowledgments} 

The research of the first author is funded by a GENES doctoral scholarship.
The research of the second author is partially supported by a grant from the
French National  Research  Agency (ANR)  as  part  of  the  Investissements
d’Avenir program (ANR-11-LABEX-0047).  We are thankful to Mathieu Gerber, two
anonymous referees and the associate editor who made comments that helped us to
improve the paper.

\bibliography{library_cleanded}

\section{Appendix}

\subsection{Proofs of main results}
\subsubsection{Proof of Corollary \ref{corl:asymptotic_variance_reduction}}
For the mixed sequences we have 
\begin{eqnarray*}
\lim_{N \rightarrow \infty} \frac{1}{N}\sum_{k=1}^N \sigma_{k,\text{qmc-mixed}}^2 = C_{\text{qmc-mixed}}^2
\end{eqnarray*}
and for the Monte Carlo estimate we have 
\begin{eqnarray*}
\lim_{N \rightarrow \infty} \frac{1}{N}\sum_{k=1}^N \sigma_{k,\text{mc}}^2 = C_{\text{mc}}^2
\end{eqnarray*}
where 
\begin{eqnarray*}
  C_{\text{qmc-mixed}}^2 = \int_{[0,1]^{s}} f(x)f(x)^T \dd x - \int_{[0,1]^{d}}\left( \int_{[0,1]^{s-d}} f(u) \dd X^{d+1:s} \right) \left( \int_{[0,1]^{s-d}} f(u) \dd X^{d+1:s} \right)^T \dd q^{1:d}
\end{eqnarray*}
and 
\begin{eqnarray*}
  C_{\text{mc}}^2 = \int_{[0,1]^{s}} f(x)f(x)^T \dd x - 
  \left(  \int_{[0,1]^{s}} f(x) \dd x \right) \left( \int_{[0,1]^{s}} f(x) \dd x \right)^T.
\end{eqnarray*}
We must show that 
\begin{eqnarray*}
    \left(  \int_{[0,1]^{s}} f(x) \dd x \right) \left( \int_{[0,1]^{s}} f(x) \dd x \right)^T \preceq \int_{[0,1]^{d}}\left( \int_{[0,1]^{s-d}} f(u) \dd X^{d+1:s} \right) \left( \int_{[0,1]^{s-d}} f(u) \dd X^{d+1:s} \right)^T \dd q^{1:d},
\end{eqnarray*}
in the sense of positive definite matrices. 
This inequality holds in the univariate case due to the Cauchy-Schwartz
inequality. In the multivariate case, let 
$\int_{[0,1]^{s-d}} f(u^k) \dd X^{d+1:s} = A\left(q^{1:d}\right)$. We rewrite:
\begin{eqnarray*}
    \int_{[0,1]^{d}} A\left(q^{1:d}\right) \dd q^{1:d} \int_{[0,1]^{d}} A\left(q^{1:d}\right)^T \dd q^{1:d} \preceq \int_{[0,1]^{d}} A\left(q^{1:d}\right) A\left(q^{1:d}\right)^T \dd q^{1:d}.
\end{eqnarray*}
In order to check the positive definiteness let $v \in \mathbb{R}^s$. We check 
\begin{eqnarray*}
    v^T \int_{[0,1]^{d}} A\left(q^{1:d}\right) \dd q^{1:d} \int_{[0,1]^{d}} A\left(q^{1:d}\right)^T \dd q^{1:d} v &\leq& v^T \int_{[0,1]^{d}} A\left(q^{1:d}\right) A\left(q^{1:d}\right)^T \dd q^{1:d} v, \\
     \int_{[0,1]^{d}} v^T A\left(q^{1:d}\right) \dd q^{1:d} \int_{[0,1]^{d}} A\left(q^{1:d}\right)^T v \dd q^{1:d}  &\leq&  \int_{[0,1]^{d}} v^T A\left(q^{1:d}\right) A\left(q^{1:d}\right)^T v \dd q^{1:d}. 
\end{eqnarray*}
While noting that $ v^T A\left(q^{1:d}\right) \in \mathbb{R}$ and $ A\left(q^{1:d}\right)^T v \in \mathbb{R}, \forall v \in \mathbb{R}^s$ we are back in the univariate case and the inequality holds. $\QED$

\subsubsection{Proof of Theorem \ref{theorem:mvt_clt_mixed_rqmc}}
The statement of the theorem 
is equivalent to $ \lim_{N\rightarrow\infty}  |\mathbb{P}(T_N \leq t) 
- \mathbb{P}(Z \leq t ) | = 0$ for all $t \in \mathbb{R}^s$, 
$T_N = N^{1/2} S_N^{RQMC}$, and 
$Z$ a random variable distributed according to the Gaussian limit. 

When conditioning on the random element $V$ in the RQMC sequence, we have that
\[\lim_{N\rightarrow\infty} \mathbb{P}(T_N\leq t|V=v) 
	= \mathbb{P}(Z \leq t)
\] for almost all $v$, by Theorem \ref{theorem:mvt_clt_mixed}, as a RQMC
sequence is a QMC
sequence with probability one.  Furthermore, 
$|\mathbb{P}(T_N \leq t|V=v)| \leq 1$, thus the function is dominated.
For all $N$ we have
\begin{align*} 
\left| \mathbb{P}(T_N \leq t) - \mathbb{P}(Z \leq t) \right|  
&  =  \left| \int_{\mathcal{B}} \left\{\mathbb{P}(T_N \leq t|V=v)  
		- \mathbb{P}(Z \leq t)  \right\} \dd \mathbb{P}(v) \right|,  \\
&  \leq  \int_{\mathcal{B}} \left|\mathbb{P}(T_N \leq t|b)  
		- \mathbb{P}(Z \leq t) \right|  \dd \mathbb{P}(v).
\end{align*}
And
\[
\lim_{N\rightarrow\infty} \int_{\mathcal{B}} 
\left|\mathbb{P}(T_N \leq t|V=v)  
- \mathbb{P}(Z \leq t) \right|  d\mathbb{P}(v) = 0,
\]
due to the dominated convergence theorem. Therefore 
\[
\lim_{N\rightarrow\infty} 
\left| \mathbb{P}(T_N \leq t) - \mathbb{P}(Z \leq t) \right|
= 0.
\] 
$\QED$

%
%
\subsubsection{Proof of Proposition \ref{theorem:decomp_variance_norm_constant_qmc}}

Since the $\thetavec_n$'s are deterministic, 
\begin{align*}
	\E{ \Zest } & = \frac 1 N \sum_{n=1}^N \frac{p(\thetavec_n)}{q(\thetavec_n)}
	\pballn 
	= \frac 1 N \sum_{n=1}^N f(\thetavec_n) 
	\\
	\Var[ \Zest ] & = \frac{1}{M N^2} \sum_{n=1}^N 
	\left\{ \frac{p(\thetavec_n)}{q(\thetavec_n)}\right\}^2
	\pballn \left\{ 1 - \pballn \right\}
\end{align*}
and 
$\left| \E{\Zest} - Z_\epsilon \right| = \OO(N^{\tau-1})$ for any $\tau>0$, by 
Koksma-Hlawka inequality. By the standard decomposition of the mean square error: 
\[ 
	\E{(\Zest - Z_\epsilon)^2} = 
	\left( \E{\Zest} - Z_\epsilon \right)^2 
	+ \Var\left[ \Zest \right] 
\]
and since $p(\thetavec_n)/q(\thetavec_n) \leq C$, 
we see that that the MSE times $M$ is $\OO(N^{-1})$.

\subsubsection{Proof of Theorem \ref{theorem:clt_importance_sampling}}

One has: 
\begin{align*}
	\left( \phiest - \EE_{p_\epsilon} \phi \right) & = 
	\left( \frac{\sum_{n=1}^N w_n \phi(\thetavec_n)}{\sum_{n=1}^N w_n}
	- \EE_{p_\epsilon} \phi \right)  \\ 
	& = \frac{N^{-1} \sum_{n=1}^N w_n \bar{\phi}(\thetavec_n)}
	{N^{-1} \sum_{n=1}^N w_n} 
\end{align*}
where $\bar{\phi} = \phi - \EE_{p_\epsilon} \phi$. 
Since the denominator converges almost surely to $Z_\epsilon$, and 
the numerator (times $N^{1/2}$) converges to a Gaussian limit (per 
Theorem \ref{theorem:mvt_clt_mixed}), we may apply Slutsky's theorem 
to obtain the desired result.  

More precisely,  the numerator has a null expectation, and is such that 
\[ 
	N^{-1/2} \sum_{n=1}^N w_n \bar{\phi}(\thetavec_n) 
	\cvl \mathcal{N}\left(0, \tau^2(\phi) \right) 
\]
where 
\[
	\tau^2(\phi) = \int_\Theta \frac{p(\thetavec)^2}{q(\thetavec)}
	\bar{\phi}(\thetavec)^2 
	\frac{b(\thetavec_n) \{1-b(\thetavec_n)\}}{M}
	\dd\thetavec
\]
again by direct application of Theorem \ref{theorem:mvt_clt_mixed}, and 
using the fact that, for a fixed $\thetavec_n$, 
\[\Var_{x_n\sim q_{\thetavec_n}} 
	[\hat{L}_\epsilon (x_n) ] = 
	\frac{b(\thetavec_n) \{1-b(\thetavec_n)\}}{M}
\]
with $b(\thetavec)=\pball$. 
$\QED$

\end{document}